 \documentclass [12pt,a4paper,     ]{report}

\usepackage{graphics}
\usepackage{times}

\DeclareFontFamily{OT1}{times}{}
\DeclareFontShape {OT1}{times}{m }{n }{ <-> ptmr }{}
\DeclareFontShape {OT1}{times}{bx}{n }{ <-> ptmb }{}
\DeclareFontShape {OT1}{times}{m }{it}{ <-> ptmri}{}
\DeclareFontShape {OT1}{times}{bx}{it}{ <-> ptmbi}{}
\usepackage{amsmath}
\usepackage{amsfonts}
\usepackage{amssymb}
\usepackage{latexsym}

\begin{document}

\title{\bf  \vspace{-3cm}  {\huge ITER:} \\
                             ~\\
                  {\it The International Thermonuclear}\\
               {\it  Experimental Reactor}\\
                             ~\\
                      {\it  and the}\\
                             ~\\
           Nuclear Weapons Proliferation Implications of
%          ---------------------------------------------
                Thermonuclear-Fusion Energy Systems
%               -----------------------------------
}

\author{~\\{\bf André Gsponer and Jean-Pierre Hurni}\\
\emph{Independent Scientific Research Institute}\\ 
\emph{ Box 30, CH-1211 Geneva-12, Switzerland}\\
e-mail: isri@vtx.ch\\}

\date{ISRI-04-01.17~~~ \today}

\maketitle

\newpage
~\\

\begin{center}
\medskip
{\Huge {\bf Abstract}}
% --------------------

~\\

\end{center}
This report contains two parts:

(1) A list of ``points'' highlighting the strategic-political and military-technical reasons and implications of the very probable siting of ITER (the International Thermonuclear Experimental Reactor) in Japan, which should be confirmed sometimes in early 2004.

(2) A technical analysis of the nuclear weapons proliferation implications of inertial- and magnetic-confinement fusion systems substantiating the technical points highlighted in the first part, and showing that while full access to the physics of thermonuclear weapons is the main implication of ICF, full access to large-scale tritium technology is the main proliferation impact of MCF.

The conclusion of the report is that siting ITER in a country such as Japan, which already has a large separated-plutonium stockpile, and an ambitious laser-driven ICF program (comparable in size and quality to those of the United States or France), will considerably increase its latent (or virtual) nuclear weapons proliferation status, and foster further nuclear proliferation throughout the world. 

%The safety and environmental problems related to the operation of large-scale fusion facilities such as ITER (which contain massive amounts of hazardous and/or radioactive materials such as tritium, lithium, and beryllium, as well as neutron-activated structural materials) are not addressed in this report.

\newpage
~\\

\begin{center}
\medskip
{\Huge {\bf About the authors}}
% --------------------------------------

~\\

\end{center}

Dr.\ André Gsponer received his PhD in physics in 1978 for an experiment performed at the Fermi National Accelerator Laboratory as a research-associate at the University of Chicago.  Since 1978 he participated in numerous international conferences dealing with fission-, fusion-, and accelerator-based nuclear energy systems.  He published many papers on the nuclear weapons proliferation aspects related to these systems, and contributed to several hearings on them, especially at the French and European parliaments.  Since 1982 he is directing ISRI, an independent scientific research institute specialized in advanced science and technology, and their implications for current and future nuclear weapons.

Dr.\ Jean-Pierre Hurni received his PhD in physics in 1987 from the University of Geneva for a research in theoretical physics.  Since the mid-1980s he is a senior researcher at ISRI where he has contributed to many projects and papers dealing both with pure science and with the technological impact of advanced nuclear science on the proliferation of nuclear weapons.

\newpage
~\\

\begin{center}
\medskip
{\Huge {\bf Foreword}}
% --------------------------------------

~\\

\end{center}

\begin{citation}

{\it ``The atomic weight of hydrogen is not exactly 1,
but by careful measurement is found to be 1.0077.
Who could imagine that in this slight discrepancy
--- which indeed needs some explanation to make
intelligible --- an immense store of possible
energy is indicated, which some day, when we have
learned how, may become accessible for good or
ill to the human race ? [...]  And if ever the
human race get hold of a means of tapping even a 
small fraction of the energy contained in the atoms
of their own planet, the consequences will be 
beneficent or destructive according to the state 
of civilization at that time attained.''}

\end{citation}

~\\

\hspace{6cm} Sir Oliver Lodge, F.R.S.

\hspace{6cm} \emph{Scientific American}, May 1924

%-------------------------------------------------

\tableofcontents

\listoftables

\listoffigures

\chapter{ITER Siting Issues}
%===========================

\noindent What follows is a collection of strategic-political and military-technical points that should be addressed in relation to the siting and construction of an experimental fusion reactor such as the \emph{International Thermonuclear Experimental Reactor} (ITER),  as well as more specific points to be considered in the case of Japan which is the most likely country in which ITER will be sited.

This list is focused on the nuclear weapons proliferation issues and implications. It does not address the impact of ITER's siting on international and national fusion energy research policies.  Neither does it attempt to give a comprehensive view of the economic and political motivations of particular States in supporting a given site. (We refer to the Appendix for a concise discussion of some of these issues in the perspective of ITER's siting in Japan.\footnote{We are indebt to Professor Clifford E.\ Singer for this discussion and other helpful comments on an earlier version of this report.})

Moreover, the points do not address the safety and environmental risks associated with the operation of thermonuclear fusion facilities, which mainly come from their use of massive amounts of hazardous materials such as tritium, lithium, and beryllium: Accidental tritium dispersal, lithium fires, toxicity of beryllium, etc. Similarly, it does not address the problems related to the production and safe disposal of the large amounts of long-lived low-radioactive waste generated by the activation of structural materials bombarded  by the neutrons produced in the fusion reactions \cite{ITEROFF,ITERCAN}.

The points have been classified according to their main emphasis, leading to a certain amount of repetition since there is considerable overlap between different issues. For every one of these points references will be given, either to following sections of this report, or to technical papers in which more details are given.  Concerning the political aspects, apart from a few specific references, we refer the reader to the vast literature, and to the numerous newspapers articles and analyses, in which many of the political-diplomatic points mentioned here are discussed by political scientists or diplomats, rather than by physicists like the authors of this report.

%\newpage 

\section{Mostly political points}
%----------------------------

\begin{itemize}

\item  The January 30, 2003, decision of the United States to return to the ITER project, almost five years after having withdrawn from it, was primarily the result of the Office of Fusion Energy and its advisory panels desire to get back into international thermonuclear plasma work seen as justifying the whole fusion energy research effort, and was made possible by the generally pro-nuclear current U.S. administration and less anti-nuclear Congress.  However, it can be argued that this decision was also facilitated by political considerations related to the ``new nuclear arms-control environment'' created by the Indian and Pakistani nuclear test of 1998 and the numerous more recent events which culminated with the North Korean claim of having assembled a number of nuclear weapons (see \cite{SINGE01} and Sec.2.1).\footnote{Other political (and possibly financial) considerations must have included the commitments of Canada and China to ITER, along with an expression of interest from South Korea. Since Canada withdrew in December 2003, the current participants in negotiations on ITER construction are China, the European Union, Japan, the Russian Federation, South Korea, and the USA.}  Since the weapons now in the arsenals of these countries are deliverable by medium to long-range missiles, they are most certainly ``boosted'' fission bombs, meaning that they contain besides some fissile material (a few kilograms of plutonium or enriched uranium) a small amount of fusion material (a few grams of tritium) to make them sufficiently compact, safe, and reliable to enable military deployment (See Fig.1 and Sec.2.2.1).  For this reason, any international enterprise such as the construction of ITER, in which kilogram amounts of tritium are used, becomes a sensitive undertaking to which the U.S.\ have to participate in order to have as much influence as possible --- especially since the current administration in Washington has taken a much stronger stand on the strengthening of the U.S.\ nuclear arsenal than on arms-control and disarmament.\footnote{Some of the post-1998 and post-September-11 decisions of the U.S.\ government which have significantly contributed in complicating the ``new nuclear arms-control environment'' are: The non-ratification of the CTBT, the withdrawal from the ABM treaty, the deployment of a national missile defense system, the possible modification of existing nuclear weapons into ``earth penetrators'' or ``bunker-buster,'' etc.}

\item Experimental fusion reactors such as ITER, as well the deployment of commercial-scale fusion reactors, pose the problem of tritium proliferation because their operation will imply the yearly use of kilograms of tritium (See \cite[Sec.6.2.5, p.28]{ITER-basis-6} and Sec.2.5.2), i.e., amounts comparable to those used in an arsenal of several thousands ``tritium-boosted'' nuclear weapons.  For example, the total amount of tritium in the U.S.\ nuclear stockpile is on the order of 100 kg, which corresponds to an average on the order of 10 g per nuclear warhead.\footnote{The National Academy of Science report of December 20, 2002, recommending that the U.S.\ enter ITER negotiations and pursue an appropriate level of involvement in ITER, strongly emphasizes the importance of ITER for demonstrating full-scale tritium burning and breeding technology \cite[p.5-7]{NRCBP02}. However, the word ``proliferation'' does not appear at all in the report.  Moreover, while concerns about ``proliferation'' are  mentioned as a minor item in some official documents such as the \emph{ITER Technical Basis} report \cite[Sec.5.2.4.1, p.6]{ITER-basis-5}, they are generally described in positive terms emphasizing that a \emph{pure} fusion reactor such as ITER will not contain \emph{fission} materials. (For a discussion of \emph{hybrid} fusion-fission reactors see Sec.2.5.1.)  Similarly, the reports of the \emph{Special Committee on the ITER Project of the Japanese Atomic Energy Commission} highlight only the ``reduced chances for nuclear weapons proliferation'' in comparison with fission reactors  \cite[p.266-267]{RSC-ITER}: The problems with tritium proliferation and the close links between fusion energy and thermonuclear weapons are never mentioned.}

\item The present stage of the ITER siting debate, in which a French site is in competition with a Japanese site, reflects the current strategic-political background: The European Union, Russia, and China are in favor of the French site of Cadarache --- which would locate ITER in a NPT-recognized nuclear-weapon State; while the United States, South Korea, and Japan are in favor of the Japanese site of Rokkasho \cite{JASS-2002} --- which would locate ITER in a non-nuclear-weapon State. An important reason for this split is that the construction and operation of ITER will give the host country full access to large-scale tritium technology --- which is presently only available to nuclear-weapon States and to Canada.  Indeed, once in full operation,  ITER's site inventory of tritium will be about 2 kg, and ITER's tritium consumption about 1.2  kg per year \cite[ Table 6.2.5-1, p.29]{ITER-basis-6}.\footnote{For the record we mention that financial and other political considerations also play an important role in the ITER siting debate.  For instance, Japan is ready to pay 48\% of construction costs and 42\% of operating cost.  But part of the \$30 billions total cost of ITER could compensate for the suppression of the Iraqi dept to Japan, as well as for the cost of sending Japanese troops to Iraq.  Moreover, some French political leaders consider that ``the American choice is explained by a will to `punish' France for its attitude during the Iraqi crisis'' \cite{AFP--2004}.}

\item As a non-nuclear-weapon State and active proponent of nuclear disarmament, Japan is particularly sensitive to the new arms-control environment created by the testing of weaponized nuclear devices by India and Pakistan in 1998,  and by the more recent North Korean claim of having assembled a number of nuclear weapons.  For instance ``Japan has been particularly adamant against the expanding of the number of states with nuclear weapons `status' beyond the current (P-5) permanent veto members of the UN Security Council'' \cite[p.16]{SINGE01}.  On the other hand, as a counterweight to this new environment, Japan is likely to welcome the enhancement of the already massive \emph{latent}\footnote{For a discussion of the concept of ``latent'' (or ``virtual'') nuclear weapons proliferation, which refers to the capability to manufacture nuclear weapons components and to assemble them on short notice, see sections 2.1 and 2.9 of this report.} nuclear capability (provided by its fission reactor and reprocessing facilities) resulting from the siting of ITER in Japan --- which will give full access to large-scale tritium technology, and which in time may require a large-scale indigenous tritium production capability, of a size comparable to one sufficient to maintain a large thermonuclear arsenal.

\item A fusion reactor program (in which every operating commercial-scale reactor requires an inventory of at least 10 kg of tritium, see Sec.2.5.2) gives a non-military justification to acquire an industrial-scale tritium technology --- just like a fast-breeder reactor program (in which every reactor requires an inventory of at least 1000 kg of plutonium) gave a civilian justification to the industrial scale reprocessing of spent nuclear fuel in countries such as Japan.  With the construction of a large fusion reactor, large-scale tritium production in a non-nuclear-weapon State becomes ``respectable,'' even though it is providing at the same time a status of latent (or virtual) \emph{advanced} nuclear-weapon-State.

\item The nuclear weapons proliferations implications of the siting of ITER in Japan is enhanced by the fact that Japan is already in possession of a large stockpile of separated plutonium.  Indeed, large-scale availability of tritium would enable this reactor-grade plutonium to be used in highly efficient and reliable nuclear explosives ``boosted'' by a few grams of tritium each (see Fig.2 and Sec.2.2.1).   Such boosted bombs are sufficiently compact to be used as warheads in relatively unsophisticated long-range missiles, or as primaries in two-stage thermonuclear bombs --- in which reactor-grade plutonium can also be used in a third-stage to enhance the yield (see Fig.4 and Sec.2.2.2).

\item Considering that Japan already has a large inertial confinement fusion (ICF) program comparable in ambition to those the United-States and France (see Tables 2 and 3, and Sec.2.4), as well as an ambitious magnetic confinement fusion (MCF) program,\footnote{See reference \cite{SUEMA03} for an extensive discussion of past, present, and future ICF and MCF research under the Japanese Ministry of Education, Sports, Culture, Science and Technology.} the access to large-scale tritium-handling, and progressively to large-scale tritium-breeding, capabilities will put Japan in a position similar to advanced nuclear-weapon-states with regards to the making of very advanced nuclear weapons, as well as of fourth-generation nuclear weapons in which fusion materials (i.e., deuterium, tritium, and lithium) will be used instead of fissile materials as the main explosive.

\item Like India and Pakistan, South Korea has already a militarily significant indigenous tritium production capacity.\footnote{The devices tested by India and Pakistan in 1998 were boosted nuclear explosives. In North Korea, the 5 MW(electric), 20--30 MW(thermal), experimental reactor at Yongbyon can produce several kilograms of plutonium every year, or several tens of grams of tritium per year --- enough to maintain a small arsenal of half a dozen boosted fission bombs.}  This is because some tritium is produced in the heavy-water of its four Canadian-built CANDU-type power reactors, which also yield significant amounts of high-grade plutonium in their spent fuel.  Therefore, South Korea's participation to the ITER collaboration will further enhance its latent nuclear-weapons capability, especially if ITER is built in Japan.  Consequently, it will become less likely that North Korea will dismantel its nuclear weapons, which may encourage further countries to develop an actual nuclear-weapons capability.

\item The Japanese proposal of siting ITER at Rokkasho is related to its commitment to complete and operate a controversial reprocessing plant under construction at this site (see Appendix). This raises the possibility that ITER could be built at Rokkasho instead of the reprocessing plant, which could be seen as a more favorable option from the point of view of nonproliferation.  However, in the opinion of the authors of the present report, this is likely not to be the case because Japan already has a large stockpile of separated plutonium (so that reprocessing more fuel would not dramatically change the situation with respect to access to fissonable material) while ITER will enable Japan to go further in the direction of latent proliferation, and closer to next-generation nuclear weapons which are likely to be of a pure-fusion type.

\item Since large-scale tritium technology is the only major thermonuclear weapon related technology not yet available to Japan, the siting of ITER in Japan may encouraged further smaller and larger countries around the world, including highly industrialized countries such as Germany,\footnote{Germany developed an indigenous tritium capability (comparable in scale to that of Japan) which led to the sale of tritium extraction equipment to Pakistan in the 1980s \cite[p.195]{MULLE92}.} to enhance their latent nuclear weapons capability.  A major responsibility of the countries promoting ITER's construction in a non-nuclear-weapon State such as Japan is therefore the additional impetus that is given to the nuclear arms race.

\item Finally, we stress that latent nuclear weapons proliferation, and its perceived value as a substitute for an actual nuclear weapons capability (see Sec.2.9), are fully compatible with the letter of the Nuclear Non-Proliferation Treaty (NPT) and of the Comprehensive Nuclear Test Ban Treaty (CTBT), despite that modern technology makes that the differences between advanced civilian-nuclear-powers and actual nuclear-weapon-States become less and less significant (see Sec.2.10).  In this context, visibility and transparency are essential to \emph{prove} the latent capability, which is enhanced by having a large thermonuclear facility such as ITER on national territory rather than elsewhere.%\footnote{From a broader strategic perspective, siting ITER at Rokkasho would also provide Japan with a kind of a ``World-Lab,'' something comparable to CERN in Europe, or to the International Space Station, which would make Japan the focal point of international thermonuclear fusion energy R\&D.}

\item In summary, building ITER at the Rokkasho site will turn Japan into a \emph{virtual thermonuclear superpower}. This will not be the result of a few deliberate political decisions, but the consequence of many independent small steps, which starting in the late 1940s \emph{institutionalized} nuclear and thermonuclear energy research in all countries, and turned them into powerful instruments of foreign policy, because of their close connections with nuclear and thermonuclear weapons (see Sec.2.9).  Like the dropping of an atomic bomb, or the starting of a war of aggression,  the next step  --- going from the virtual to the actual state of proliferation --- may then depend much less on a democratic decision process than on the availability of the technology, which is why the ITER siting decision should not be taken without a thorough parliamentary assessment of all technical, political, and military implications.

\end{itemize}

\section{Problems related to tritium transportation, diversion, and theft}
%---------------------------------------------------------------------

\begin{itemize}

\item The transportation of large quantities of tritium to Japan is a new and particularly sensitive issue because at present most of the tritium never leaves the national boundaries of nuclear-weapon States.  The present-day industrial and scientific uses of tritium are such that only relatively small amounts are shipped between countries.  On the other hand, ``it is expected that about 25 kg of tritium will be transported during the course of ITER operations'' \cite[p.18]{JASS-2002}.  The \emph{Japan Atomic Energy Research Institute} (JAERI) ``has some previous experience with tritium transportation from Canada (3 times) and the USA (6 times).  It is estimated for the deuterium-tritium operation in ITER that approximately 150 g of tritium is transported six times a year by using a transport package with a 50 g-tritium capacity each'' \cite[p.18]{JASS-2002}.\footnote{For comparison, the current world tritium market corresponds to the shipment of about 100 g/year, mainly from Canada, and in very small batches.}  

\item Until now, all thermonuclear fusion facilities in which substantial amounts of tritium are used were always built closed to a nuclear weapons facility or research laboratory.  This was the case, for example, of the \emph{Joint European Torus} (JET) which was built at Culham, i.e., near Aldermaston, the main U.K.\ nuclear weapons research laboratory.  In the case of the Japanese site of Rokkasho, it can be argued that a U.S.\ military base is located nearby.\footnote{According to the International Herald Tribune: ``The Japanese site has many assets: the proximity of a port, a ground of solid bedrock and the close proximity of a US military base'' \cite{IHT--04}.}  However, if tritium transportation is made using this base instead of a civilian entry point to Japan, national and international political problems are likely to result.

\item Considering the political, social, and environmental problems already encountered with the transportation of spent nuclear fuel and separated plutonium between Europe and Japan, the transportation of large amounts of tritium between Canada and Japan will add a new dimension to the risks associated with nuclear energy.  For this reason it is likely that once the ITER siting decision will be taken, discussion will start on the construction of a dedicated tritium production facility in Japan, e.g., possibly a system based on the use of a high-intensity accelerator of the kind which is already under development in Japan (see end of Sec.2.5.3).  

\item However, while the construction of a tritium production facility close to ITER will decrease the need for transportation over long distancies, it will not completely suppress the possibility of tritium diversion or theft.\footnote{We do not suggest that tritium could be diverted into a clandestine Japanese weapons program: As long as a latent proliferation status is implicitly perceived as giving sufficient guarantees for developing an actual capability in case of need, there is no reason for a weapons program.}

\item The concealment of a military significant amount of tritium after a diversion or theft is much easier than that of a comparable amount of plutonium.  This is because a militarily significant amount of tritium is comparatively much smaller than a militarily significant amount of plutonium (see Sec.2.5.2), and because tritium is much lighter, much less radioactive, and more difficult to detect than plutonium.

\item The problems of tritium diversion or theft are exacerbated by the fact that  small amounts of tritium can be used to considerably enhance the yield of crude nuclear weapons that could be assembled by terrorists or minor rogue states using low-grade materials such as reactor-grade plutonium or low-enriched uranium.

\end{itemize}

\section{Technical points related to thermonuclear weapons and  their proliferation}
%---------------------------------------------------------------------------

\begin{itemize}

\item A few grams of tritium are sufficient to ``boost'' fission explosives made of a few kilograms of military- or reactor-grade plutonium.  (As is explained in section 2.1, this is because tritium boosting obviates the preignition problems which makes non-boosted plutonium weapons unreliable and unsafe).  Such boosted explosives are sufficiently compact to be used in relatively unsophisticated long-range missiles, or as primaries of two-stage thermonuclear bombs (see Fig.4 and Sec.2.2.2).

\item A few tens of grams of tritium are sufficient to ``boost'' the ``sparkplug'' of the secondary of a two-stage thermonuclear weapon, which can then be made sufficiently small to be deliverable by a missile.  In such weapons a ``third-stage'' made of a few tens of kilograms of fissile material is often used to provide additional yield without increasing the size and weight of the warhead. While this third-stage (i.e., essentially a ``blanket'' surrounding the secondary of a two-stage H-bomb) is preferably made of enriched uranium, it can also be made of reactor-grade plutonium without degrading the military performance of the device, which may correspond to yields in the range of 200--500 kilotons equivalent TNT (see Fig.4 and Sec.2.2.2).

\item Next-generation (i.e., fourth-generation) nuclear weapons will have substantially lower yields (i.e., tons instead of kilotons) and will rely on fusion materials rather than on fissile materials for their main explosive charge (see Sec.2.6.3).  This means that the emphasis of safeguarding militarily useful materials will gradually shift from fission materials such as plutonium to fusion materials such as tritium.

\item Both magnetic confinement fusion (MCF) reactors, such as ITER, and inertial confinement fusion reactors (ICF), such as the GEKKO XII laser facility  of the \emph{Institute of Laser Engineering} at Osaka University in Japan, have nuclear-weapon-\emph{materials} proliferation implications due to their use of tritium (see Sec.2.5).  However, ICF facilities have the additional proliferation problem that they enable the \emph{physics} of thermonuclear weapons to be studied in the laboratory (see Sec.2.6).  As a result, it is well known by nuclear proliferation experts that Japan has not only the capability to build boosted nuclear weapons, but also the potential to build two-stage thermonuclear weapons that are likely to work first-time without testing.\footnote{Quoting from a U.S. Department of Energy Office of Arms Control and Nonproliferation study of 1995, ``one cannot rule out that a technologically advanced country would be able to field a very conservatively designed thermonuclear weapon that would present a credible threat without nuclear testing'' (see Sec.2.1 and Sec.2.2.1).}  Today, the main impediment that would prevent Japan from building such second-generation nuclear weapons on short notice is the unavailability of sufficient amounts of tritium.

\item Any future commercial fusion reactor (based on the either the MCF or ICF principle) poses the problem of tritium proliferation because during operation each such reactor will contain several tens of kilograms of tritium, i.e., enough for an arsenal of several hundreds or thousands thermonuclear weapons (see Sec.2.5.2).

\item Since 1 g of deuterium-tritium fuel produces about 340 GJ of energy, a nominal 1 GW (electric) commercial fusion power plant with a thermal efficient of 30\% would consume 10 mg of $DT$-fuel per second.\footnote{This corresponds to a consumption of 0.5 kg of tritium per day.} If the plant is based on the MCF principle this amount of fuel is burnt relatively slowly in a steady-state plasma confined by magnetic fields. However, if the plant is based on the ICF principle, the fuel is burnt in a continuous salvo of micro-thermonuclear-bombs (so-called ``pellets'') exploding at the center of a containment vessel.  Assuming that one pellet is detonated each second, the explosive yield of each pellet would be 3.4 GJ, i.e., equivalent to about 810 kg of TNT (see Sec.2.4).

\item The many kilograms of tritium implied in the daily operation of a number of GW-scale reactor in a fusion-based power industry will provide numerous possibilities of diversion or theft of tritium for military or terrorist purposes.  With ICF reactors, one will have the additional problem of safeguarding the fusion pellets themselves, because they may possibly be used as the explosive charge of some future-generation nuclear weapons.

\end{itemize}

\section{Japan's expertise with tritium technology}
%----------------------------------------------

\begin{itemize}

\item Japan's scientific and technical experience with tritium goes back to 1956 with the delivery of the first supply of tritium from England to the \emph{Tritium Research Center} (now called \emph{Hydrogen Isotope Research Center}) at Toyama University.

\item Technical experience with gram amounts of tritium is currently available at several Japanese laboratories, in particular at the \emph{Japan Atomic Energy Research Institute} (JAERI) at Tokai, where the technology for mass-producing lithium-ceramic and beryllium pebbles to be used in tritium breeding blankets of fusion reactors is being developed in collaboration with Japanese industry \cite{KAWAM00}.  These pebbles are likely to be used in ITER tritium breeding experiments.  But they could also be used in some future indigenous tritium production facility.  Further experience related to the use of tritium in ICF pellets exists at the \emph{Institute of Laser Engineering} at Osaka University.

\item Since 1994 JAERI and U.S.\ scientists are collaborating in research and development related to tritium production in a fusion blanket. This has lead to actual tritium fuel processing experiments using the TSTA (Tritium Systems Test Assembly) at the Los Alamos National Laboratory \cite{KONIS95}.  TSTA is the only existing simulated loop of the fusion processing cycle of a fusion reactor capable of handling about 100 grams of tritium.

\item According to current plans described in the \emph{ITER Technical Basis}  it is expected that ITER will be operated with all tritium supplied by external sources.  ``The maximum total tritium transportation per year will be about 1.2 kg/year in the first ten years.  Assuming a 50 g tritium transport container, there will be two shipments every month''  \cite[Sec.6.2.5, p.28]{ITER-basis-6}.  ``A reasonable estimate of the available tritium from Canada [where tritium is bred as a byproduct of the operation of twenty CANDU-type reactors], for example, is: 22 kg in storage by 2009; 1.5 kg/year from 2009'' \cite[Sec.6.2.5, p.28]{ITER-basis-6}. 

\item  It is expected that ``the CANDU tritium supply [from Canada] will be available for use in ITER in its initial phase of operation, but will not enable ITER to run its extended phase of operation at reasonable device availability'' \cite{ABDOU03}.  However, ``typical tritium production capacity from fission reactors specially designed for tritium production is only a few kg per year, and at the prohibitive cost of about \$200 million dollars per kg''  \cite{ABDOU03}.  This calls for new tritium production facilities, e.g., accelerator or dedicated fusion-based tritium breeders.  Moreover, according to a Los Alamos study on tritium supply for ITER : ``Development and deployment of program components which breed significant quantities of tritium are needed soon'' \cite{WILLM03}.

\item Once the supply of tritium from Canada will be exhausted (because of the foreseeable shut-down of aging CANDU reactors), or in case of a problem with transportation, the most logical source of tritium would be a dedicated facility in the host country.  This could be a justification for an accelerator breeder system, based on a high-intensity proton accelerator of the kind considered in Japan (and in South Korea) for neutron science and nuclear waste transmutation (see end of Sec.2.5.3).  To have such a production facility available in due time, development is to start soon \cite{WILLM03}.

%\item  Similarly, to avoid problems with transportation,  the most likely place for reprocessing the ITER blanket will be in the host country, especially if the JAERI tritium breeding pellets are used in the blanket and if ITER is built in Japan.

\item In summary, while Japan has now experience and facilities to handle tritium in gram amounts, it has not yet access to kilogram-amount tritium production and processing technology, nor the justification to develop it.  However, if ITER is built in Japan, the problems with transportation and medium to long-term tritium supply will provide many political and technical justifications to go ahead with a full-scale national tritium production program.

\end{itemize}

\chapter{Nuclear Weapons Proliferation Issues 
        of Thermonuclear-Fusion Energy Systems}
%==============================================

\section{Introduction}
%=====================

In the past decade there has been a growing awareness that the development of thermonuclear fusion systems is being accompanied by the spread of the knowledge and materials required for the production of thermonuclear weapons.

This awareness is the result of a number of political and technical developments which shifted part of the nuclear weapons proliferation's debate away from its traditional focus (namely the nuclear-\emph{fission} fuel cycle and its related enrichment and reprocessing technologies) and started to highlight the military and political problems associated with emerging nuclear energy systems: Thermonuclear-\emph{fusion} systems for energy production and nuclear weapons simulation, accelerator-based concepts for energy amplification and tritium breeding, pulsed-power technologies for civilian and military applications, etc.

Some of the more important political events which contributed to this shift include the U.S.\ Government \emph{declassification act of 1993} which ``legitimized'' the growing involvement of non-nuclear-weapon States in the development of inertial confinement fusion systems; the last round of French and Chinese underground nuclear tests which highlighted the importance of actual data for comparison with the output of future computer simulation programs; the negotiation and the conclusion of a Comprehensive nuclear test ban treaty (CTBT) which saw the opposition of India to the other parties because computer simulations and laboratory tests were \emph{not} included in the scope of the treaty; and the establishment of extensive and well-funded nuclear weapons simulation and ``stockpile stewardship'' programs in the United States and France which confirmed the primordial importance of large laser-fusion facilities for these purposes.

On the other hand, a surprising number of significant technological breakthroughs were made during the same period:  An increase by a factor of one million of the intensity of ultrashort-pulselength high-power tabletop lasers,  the capturing and cooling of antimatter, the discovery of high-temperature superconductors, the design of high-performance electromagnetic guns, the synthesis of super-heavy elements and the discovery of several unexpected new types of nuclear species (non-fissile shape isomers, medium-weight superdeformed nuclei, halo nuclei, etc.), the emergence of nanotechnology as a radical approach to the fabrications of new materials and micromechanisms, the development of new techniques for constructing ever more powerful supercomputer, etc.  All these technologies have direct military applications, for both non-nuclear and nuclear weapons, as well as many applications for possible next-generation nuclear weapons and various future thermonuclear energy systems. 

However, much more than these evolutionary changes, the most important nuclear-proliferation event of this last decade is certainly the May 1998 decision of India to explode five first- and second-generation nuclear devices (including a two-stage hydrogen bomb),\footnote{A plausible description of the five Indian tests is as follows: one of the tests was a ``certification test'' of a conservative weaponized design that must be part of the current Indian arsenal, the ``high-yield test'' was a crude two-stage hydrogen bomb, and  the ``low-yield tests'' were three advanced boosted devices for use as tactical weapons or H-bomb primaries.} which after two weeks of suspense and intense diplomatic activity were reciprocated by Pakistan with a series of six nuclear explosions.\footnote{There is less technical information on the six Pakistani tests.  However, according to an interview of Abdul Quader Kahn, the architect of Pakistan's nuclear program, the devices were high efficiency, highly reliable enriched uranium devices. ``One was a big bomb which had a yield of about 30--35 kilotons [...]. The other four were small, tactical weapons of low yield.'' In an other interview, he confirmed that ``the devices tested on 28 May were boosted weapons, as were some of the Indian tests.''}

Indeed, it soon became clear that if the bombs included in the tests were relatively compact devices ready for delivery by aircrafts or missiles, they were not some kind of ``crude'' second-world-war type of weapons, but fairly advanced ``tritium-boosted'' devices --- as are, most probably, those recently claimed to have been assembled by North Korea.\footnote{The 5 MW(e), 20--30 MW(th), experimental reactor at Yongbyon can produce several kilograms of plutonium every year, or several tens of grams of tritium per year --- enough to maintain a small arsenal of half a dozen boosted fission bombs.}  This highlighted the fact that thermonuclear-fusion materials such as tritium, and the related scientific/technical knowhow, are just as important as the much better known nuclear-fission materials for making deliverable nuclear weapons.  Moreover, it reminded that if countries like India, Pakistan, and North Korea were able to deploy such weapons, countries that are relatively much more advanced in science and technology should easily be able to do the same, or better, especially if they have access to advanced nuclear and thermonuclear facilities.

As a result, the current situation with nuclear weapons proliferation is in almost every way different and more complex than it was only a couple of years ago.  For instance, the term ``proliferation of nuclear weapons'' used to cover (i) the increase in the number and the quality of such weapons within the five ``official'' nuclear-weapon States (namely\footnote{Article IX of the Nonproliferation treaty of 1968 defines a nuclear-weapon States as ``one which has manufactured and exploded a nuclear weapon or other nuclear explosive device prior to 1 January 1967.''} China, France, Russia the U.K.\ and the U.S.A.); and (ii) the spread of nuclear weapons to other countries.  While the former is known as \emph{vertical} proliferation, the latter is called \emph{horizontal} proliferation.

Today, with an overall trend towards a decrease in the number of deployed nuclear weapons in the ``official'' nuclear-weapon States, the emphasis of \emph{vertical proliferation} is shifting away from the problem of more and better nuclear weapons to the question of the feasibility of \emph{modifying existing types} of nuclear weapons for new missions such as deep earth penetration, or of \emph{radically new types} of nuclear weapons, i.e., fourth-generation nuclear weapons, that could possibly be better adapted to perceived new military needs. 

Similarly, with the stabilization and the possible decrease in the number of additional countries likely to acquire first- and second-generation nuclear weapons,\footnote{E.g., in alphabetical order, Algeria, Egypt, Indonesia, Iran, Iraq, Kazakhstan, Libya, South Korea, Ukraine, Taiwan, etc.} the traditional \emph{horizontal proliferation} concern (which mainly focussed on the possible spread of fission-weapons) is shifting to the problem of the  proliferation of \emph{thermonuclear-fusion} weapons technology (and possibly of the technology of even more advanced types of nuclear weapons) to Israel, India, and Pakistan, as well as to countries which already have the technical \emph{capability} to build {nuclear-fission} weapons but which have decided not to build them.

Indeed, as we now know for a fact, during the 1950s to the 1980s, a number of larger and smaller industrialized countries\footnote{E.g., in alphabetical order, Argentina, Belgium, Brazil, Canada, Germany, Italy, Japan, Poland, South Africa, South Korea, Spain, Sweden, Switzerland, etc.} have acquired the technical and industrial \emph{capability} (e.g., by means of ``peaceful'' nuclear activities) to manufacture nuclear weapons components and to assemble them on short notice.  Because such an approach to a nuclear weapon capability is inherently ambiguous and does not force a nation to signal or even decide in advance its actual intentions, it is termed \emph{latent} proliferation \cite{FEIVE72}.\footnote{In 1993-95, the concept of \emph{virtual nuclear arsenals} has been suggested to describe a world in which all assembled, ready-for-use nuclear weapons would be banned \cite{MAZAR95}. This would link \emph{latent} proliferation to deterrence in  a way that has been termed \emph{factory deterrence} by political scientists \cite{WALTZ97}, or \emph{technical deterrence}, \emph{scientific deterrence}, or \emph{deterrence by competence}, by nuclear weapons designers \cite{YOUNG97}.  In this review, we use the adjectives ``latent'' and ``virtual'' interchangeably to refer to the capability to manufacture nuclear weapons components and to assemble them on short notice.}

Moreover, it is now also recognized that some large industrialized countries, e.g., Japan and Germany, have extended their latent proliferation capability to the point where they could make not only atomic bombs but also hydrogen bombs that could be built and delivered with a very high probability of success. In fact, quoting from the  U.S. Department of Energy Office of Arms Control and Nonproliferation, 
\begin{quote}
``one cannot rule out that a technologically advanced country would be able to field a very conservatively designed thermonuclear weapon that would present a credible threat without nuclear testing'' \cite[p.27]{USDOE95}.
\end{quote}

It is in this context of growing technical complexity and expanding political ambiguity that this review is written.  In order to reduce some of the uncertainties associated with this situation, and to make the links between thermonuclear weapons and thermonuclear fusion energy systems as clear as possible, it is therefore important to start by reviewing the main technical concepts.  This is done in section 2.2, 2.3 and 2.4 were some basic information on thermonuclear weapons and fusion energy systems is recalled and updated to fit into the contemporary debate. The proliferation implications that are common to all thermonuclear fusion energy systems will be addressed in section 2.5. Then, the more specific implications of the main fusion systems (i.e., \emph{inertial} confinement fusion, ICF, and  \emph{magnetic} confinement fusion, MCF) will be expounded in sections 2.6 and 2.7, leaving for section 2.8 those of one example of the many alternative fusion systems (AFS) that are being developed concurrently. Finally, before  concluding, section 2.9 will discuss some aspects of the problem of openness versus secrecy in matters of thermonuclear energy and thermonuclear explosives, and their implications for the latent proliferation of thermonuclear weapons.

\section{Thermonuclear weapons}
%==============================

Despite the tight secrecy that covers the technical details of how nuclear and thermonuclear weapons are built, their principles are sufficiently well known to be described fairly accurately \cite{GSPON97a}.

Using the wording of the U.S. Department of Energy, Office of declassification, the fundamental idea is that, ``in thermonuclear weapons, radiation from a fission explosive can be contained and used to transfer energy to compress and ignite a physically separate component containing thermonuclear fuel'' \cite{USDOE94}.  This is the essence of the so-called ``Teller-Ulam'' principle that was declassified by the U.S. in 1979, and which is used in all modern fusion explosives.

In practice, in order to apply this principle and build a device that is sufficiently small to be deliverable by an aircraft or a missile, it was necessary to first miniaturize the fission bomb (also called the trigger, or the primary) that is producing the radiation (i.e., soft x-rays) necessary to compress and ignite the fusion material (also called the secondary, or second-stage).  This miniaturization was achieved by using a small amount of deuterium and tritium, a thermonuclear fuel mixture, to enhance the performance of a fission bomb (See Fig.1).  This technique is called ``boosting'' because it was first developed in order to increase the yield of fission bombs. Boosting is now primarily used to decrease the overall weight and size of fission bombs of a given yield, as well as to dramatically increase their safety.  Since boosting is presently used in essentially all modern nuclear weapons (i.e., in all tactical or strategic weapons within all contemporary nuclear arsenals, including Israel, India, Pakistan, and possibly North Korea) it is important to explain this technique in some details.

\subsection{Boosted fission bombs}
%---------------------------------

Figure~2  is a simplified diagram of a boosted fission device. Its core consists of a plutonium and/or enriched uranium shell (the ``pit'') surrounded by a stainless steel case, possibly a neutron reflector, and by chemical explosives. This corresponds to the present-day concept of sealed pits, with the fissile material permanently sealed within the high explosives. A short time before detonating the device, the pit is filled with a few grams of a deuterium-tritium ($DT$) gas mixture at a pressure of a few tens of atmospheres.

When the weapon is detonated, the pit and the case are imploded by the high explosives at the same time as the  $DT$ gas. As the pit collapses into a solid ball, the  $DT$ is compressed into a sphere of a few  millimeter  radius with a density tens of times greater than its normal solid-phase density. At the same time the fissile material is compressed to a few times its normal density and the fission chain reaction starts.

As the chain reaction develops the fissile materials begins to emit x-rays and neutrons which heat the $DT$ at the center of the device.  The temperature of this mixture of fusionable materials therefore rises at the same time as the temperature of the fissile material. This leads to a remarkable phenomenon which is the cause of the boosting process: the fusion fuel ignites \emph{before} the fission chain reaction is terminated \cite{GSPON97a}.

Therefore, at a time when the diverging chain reaction has generated a yield that is still negligible, the $DT$ mixture burns out very quickly and generates a very intense pulse of high-energy neutrons by the thermonuclear reaction: $D + T  \longrightarrow  {^4\!H\!e} + n$~.  These fusion neutrons interact with the fissile material, causing it to fission, and therefore to generate most of the yield of the explosion. In other words, with boosting, the yield of a fission explosive is controlled by the very fast neutron burst from the thermonuclear reactions: the fissile material (apart from heating the fusion fuel to ignition) is essentially a passive neutron and energy amplifier in the final stage of the nuclear explosion. This leads to several important conclusions:

1)  With boosting, it is possible to build a relatively high yield fission explosive which is fairly compact because it uses only a relatively small amount of high explosives to implode the fissile material.  The device can also be made relatively  light-weight because a thick neutron reflector and/or a heavy tamper surrounding the fissile material are not necessary.

2) In an actual weapon, before arming the device, the  $DT$ mixture, or just the tritium, is stored outside of the pit in a separate reservoir. This facilitates maintenance and insures that boosting will not happen in case of an accidental detonation of the high explosives. Since the amount of high explosives needed to implode a boosted-device is only on the order of a few kilograms, a boosted fission-weapon is extremely safe because an accidental nuclear explosion is almost impossible to take place.  This increased safety is the most important single factor which enabled so many nuclear weapons to be deployed for so many year.  It is also the main reason why threshold nuclear States such as India, Israel, Pakistan, and North Korea rely on tritium-boosting technology to maintain a credible nuclear arsenal.\footnote{The gun-assembly type enriched-uranium weapons that were built by South Africa are an example of a very unsafe design. This reduces substantially the merit of the South African government for having dismanteled these weapons. In the case of Pakistan, it is unlikely that their nuclear deterrent would be based on primitive gun-assembly or implosion type weapons: besides from being unsafe, they would be much too heavy and cumbersome to be delivered by the aircrafts available in their air-force, or by their 1500 km range ``Ghauri'' missile that was tested for the first time shortly before the Indian government decided to become a declared nuclear power.  Similarly, in the case of North Korea, boosted fission devices are required for delivery by their long-range missiles.}

3) The most important technical aspects of boosting (e.g., that during the implosion of the pit by chemical explosives the fusion fuel gets sufficiently compressed without mixing with the fissile material) can be tested  \emph{without} actually starting fission or fusion reactions. This can be done outside of the scope of the CTBT, and only requires conventional equipments that are available in most high-explosive research laboratories.

4) Using boosting, it  is straightforward to build highly efficient and reliable fission weapons using \emph{reactor-grade} plutonium. In particular, the possibility of a preinitiation of the chain reaction, which creates difficulties in making a non-boosted fission bomb \cite{KANKE88, MARK93}, is no longer a serious problem.  In fact, two of the five devices tested by India in May 1998 are believed to have used plutonium that was not classified as weapons grade \cite{NATUR98}.  Moreover, independently of the type of fissile material used, the construction of ``simple'' and ``deliverable'' tritium-boosted nuclear weapons can be easier than the construction of primitive Hiroshima or Nagasaki type atomic bombs: the main problem is to acquire the few grams of tritium that are needed for every weapon (see Fig.1).

Therefore, in summary, the very important advance in fission weapons constituted by boosting \cite[p.312]{WOOD88}, and the fact that boosted bombs used as primaries are ``lower-bounding the size and mass of hydrogen bombs''  \cite[p.313]{WOOD88}, confirm the tremendous importance of tritium\footnote{And therefore of thermonuclear fusion technology.} from the point of view of the nonproliferation of fission weapons.

\subsection{Two-stage hydrogen bombs}
%-------------------------------------

Let us return to the design of a two-stage thermonuclear weapon. Its general principle, the Teller-Ulam method, was given at the beginning of this section, and is recalled in the caption of Fig.3.  As with boosting, two conditions have to be satisfied: (a) the thermonuclear fuel has to be sufficiently compressed for the fusion reaction to be fast enough, and (b) the thermonuclear fuel has to be brought to a sufficiently high temperature for the fusion fuel to be ignited. Both conditions can be satisfied by using a boosted fission bomb as a powerful source of x-rays.\footnote{This is why boosting is so important: the energy of the fission explosion (which is mostly heat in the form of x-rays) can easily radiate away from the fissile core without being attenuated by the thick neutron reflector and the large amounts of high-explosives that would be needed for a non-boosted fission bomb.}

Referring to  Fig.3, the Teller-Ulam method is as follows: a fission bomb and a container filled with fusion fuel (the secondary) are placed within a common enclosure (the radiation case); while the radiation case and the envelope of the secondary (the pusher/tamper) are made of heavy materials opaque to x-rays, the remaining space within the radiation case (the hohlraum) is filled with light-weight materials transparent to x-rays; as the primary fissions, large amounts of x-rays are radiated ahead of blast and instantaneously fill the hohlraum; x-ray radiation trapped within the hohlraum rapidly turns the hohlraum filling into a hot plasma; radiation-driven thermalization insures that this plasma has very uniform pressure and temperature so that its effects on the secondary are the same from all sides; the plasma reradiates longer wavelength x-rays that are absorbed by the surface of the secondary; the surface of the secondary (the pusher/tamper) is heated to the point where it vaporizes and material is ejected from it; the material ablated from the pusher/tamper causes by reaction a pressure which pushes it inwards, imploding the fusion fuel to very high densities. This satisfies condition (a).

Condition (b), ignition, is achieved by an optional element not yet discussed: the \emph{sparkplug} at the center of the secondary in  Fig.3. It consists of a subcritical amount of fissionable material compressed at the same time as the secondary. Because of the intense neutron background resulting from the explosion of the primary, a fission chain reaction starts in the sparkplug as soon as it becomes critical (in order to avoid a fizzle, and to maximize the yield from the sparkplug, it is generally boosted by a small amount of $DT$, on the order of 10 grams). Hence, with a careful design, the sparkplug will explode just when the thermonuclear fuel is imploded to its maximum density. It will then provide, in the form of x-rays, neutrons and additional compression from within, a large amount of energy sufficient to insure that ignition will start even in the worst case.

Once the fusion reaction starts, the fusion fuel (e.g., liquid deuterium in the first H-bomb, dry deuterated lithium {\it (}$Li^6D${\it )} in the modern H-bombs, possibly liquid deuterium-tritium in ``neutron bombs,'' etc.) will burn as long as it will be contained by the ``tamping effect'' (i.e., inertia) provided by the weight of the unablated part of the pusher/tamper. Hence the dual function of the envelope of the secondary: ``pusher'' during the ablative compression process, ``tamper'' during the thermonuclear burn and expansion of the fusion fuel.\footnote{A third possible function or the pusher/tamper, to be discussed at the end of this section, is to provide additional explosive yield.}

Consequently, when Edward Teller invented the sparkplug concept, soon after discovering with Stan Ulam in 1951 a means for achieving very high compressions, the whole scheme became thoroughly convincing. Indeed, as will be stressed much later (1983) by Carson Mark, the Los Alamos physicist who led the theoretical work on the first hydrogen bomb: ``Almost immediately (the Teller and Ulam method) gave promise of a feasible approach to thermonuclear weapons, \emph{provided only the design work be done properly}'' \cite[p.162]{MARK83}. Thus, a major feature of the Teller-Ulam design is that it provides a straightforward and intrinsically fail-safe method for making a thermonuclear bomb. In fact, this method is so good that \emph{all} the first hydrogen bombs of the five nuclear-weapon States worked the \emph{first time}.\footnote{This is why there is no doubt that any technologically advanced country can build a militarily usable hydrogen bomb without nuclear testing \cite[p.27]{USDOE95}.}  Most recently, this fact was confirmed by India: it exploded a two-stage hydrogen bomb in the series of five tests which was its second experimental campaign, 24 years after its first single experiment of 1974.\footnote{Similarly, China detonated its first full-yield hydrogen bomb after only three fission bomb tests, one boosted-fission test, and one preliminary two-stage hydrogen bomb principle test.}  According to an official Indian statement, the device worked as expected, providing a total fission-fusion yield of about 43 kilotons, a value that was intentionally kept low ``to meet stringent criteria like containment of the explosion and least possible damage to buildings and structures in the neighboring villages.''

A sketch of a modern thermonuclear bomb is given in Fig.4.  One recognizes the same elements as in Fig.3, which are found in all thermonuclear explosives, with a number of variations.  For instance, while the secondaries of early H-bombs were generally cylindrical, the most modern ones are probably of spherical shape.  In that case, instead of being made of plutonium or $U^{235}$, the sparkplug could be a $DT$ fuze which ignites the  $Li^6D$ fuel when optimum compression is reached.  Finally, by chosing for the pusher/tamper an appropriate heavy material it is possible to control the total yield of a thermonuclear weapon. This is because the high-energy (i.e., 14 MeV) neutrons produced by fusion have sufficient energy to fission any kind of transuranic material such as all isotopes of uranium and plutonium.  Therefore, while the minimum yield of a given thermonuclear design will be provided by using a non-fissionable heavy material such as lead or bismuth as the tamper, the use of depleted (i.e., mostly isotope 238), natural, or highly enriched (i.e., mostly isotope 235) uranium will provide a range of increasing yields because more and more fission reactions will be produced in the tamper.

This is why the pusher/tamper is sometimes called the ``third-stage'' of a thermonuclear weapon.  For example, assuming as in Fig.4 that the yield is $150 kt$ if the pusher/tamper is made of $U^{238}$, the yield will be  $300 kt$ if the pusher/tamper is made of $U^{235}$.  On the other hand, if the pusher/tamper is made of a  lead or bismuth, the yield will be significantly lower, on the order of $50 kt$.  This is most probably what the Indian scientists have done in 1998 in order to be able to detonate the device at a relatively low depth into the ground, and to minimize the background signals which may overload the measuring instrumentation.

Finally, if plutonium is used for the pusher/tamper a yield somewhat higher than with enriched uranium is obtained.  This is why using plutonium for the ``third stage'' is providing the highest possible yield for a given design.  This option has been implemented in some of the French thermonuclear weapons, and was an important justification for the French fast-breeder program,\footnote{In 1978, less than a year after the construction of the fast-breeder Superphenix had been undertaken, General Jean Thiry, former director of the French nuclear test site in the Pacific Ocean, advisor of the Managing  Director of the French Atomic Energy Commission (CEA), wrote: ``France is able to build atomique weapons of all types and all yields.  At relatively low cost, she will be in a position to produce large quantities of such weapons, with fast breeders providing an abundant supply of the plutonium required.'' Le Monde, 19 January 1978.} as it could be for the Indian plutonium reprocessing/recycling program \cite{NATUR98}. Consequently, although plutonium has some environmental disadvantages because it is more radioactive than $U^{235}$, it offers an attractive alternative to the large-scale production of highly enriched uranium because plutonium of \emph{any} grade can be used for the ``third stage'' of a thermonuclear weapon.  Therefore, any country such as Japan which has a large stockpile of separated plutonium will not need access to large amounts of $U^{235}$ should it decide to make high-yield thermonuclear weapons.

\section{Fusion energy systems and military technologies}
%========================================================

The main fundamental difference between fission and fusion energy systems is that the feasibility of fusion energy is still \emph{not} proven, whereas the scientific and technical feasibility of fission was established right from the beginning.  Moreover, in 1942, the construction and operation of the first ``atomic pile'' by Enrico Fermi and his collaborators proved to be so simple that it was possible to design, build and put into operation several full-scale, 1000 MW size, reactors in less than two years!  In comparison, after almost fifty years of fusion energy research, even the most optimistic scientifically-competent proponent of fusion energy would not bet to see an operating thermonuclear reactor in his life-time.

As a result, ``thermonuclear fusion'' is essentially a vast research program in which many different concepts are still competing to possibly emerge as a technically, economically, and politically viable new source of energy.  However, compared to other energy research programs, fusion research is primarily done within (or in close collaboration with) military establishments  for at least  three reasons:  First, they generally make use of high-energy density technologies which have many military applications in areas such as radars, ordnance, detonics, weapons-physics and weapons-effects simulation, etc. Second, any type of working fusion system would be a powerful source of energy, neutrons, and militarily important materials such as tritium. Third, any working fusion system that could be made sufficiently compact would be a potential pure-fusion\footnote{The concept of ``pure-fusion'' refers to the idea of burning or exploding fusion-fuel without using fission-reactions or a fission-explosive as a primary.} weapon, and could be used as an explosive or as a radiation weapon.

Therefore, when examining the possibility of ``non-electric power generation'' applications of thermonuclear fusion, one finds mostly \emph{dual-purpose civil-military applications}.  This is clearly the case with the conclusions of a 2003 report to the U.S.\ DOE Future Energy Science Advisory Committee (FESAC), which found that the most promising opportunities for non-electric applications of fusion fall into four categories \cite{MCCAR03}:

\begin{enumerate}
\item Near-Term Applications (production of medical isotopes and of neutrons for Homeland Security missions);
\item Transmutation (of nuclear waste or surplus weapons-grade plutonium);
\item Hydrogen Production (for transportation);
\item Space Propulsion (for heavy payload deep-space or manned missions).
\end{enumerate}

A further aspect which complicates the discussion of fusion energy systems is that the technologies on which they are based also have many \emph{scientific applications},\footnote{Which is why the quest for ``pure knowledge'' is often used as a justification for developing technologies which enable the study of extreme states of matter which only exist in nuclear weapons and stellar explosions.} as well as considerable overlap with those of nuclear fission energy systems. In fact, as is shown by the systematics of the numerous existing and proposed (thermo)nuclear energy systems: fission, fusion, and various high-energy beam processes, and their related technologies, are interdependent in many ways \cite{HARMS80}.\footnote{The growing interdependence of fission, fusion, acceleration, and laser technologies have also been noted by international lawyers.  It is reflected by the scope of recent export-control and technology transfert agreements such as the ``Warsaw guidlines'' and the ``Wassenaar arrangements,'' and of UN sanctions against nuclear-proliferating countries \cite{GSPON96}.}

For example, in the production of energy from fission or fusion, high-energy beam technology is important because fusion is commonly initiated  using such beams (e.g., by heating a plasma in MCF, or compressing a $DT$ pellet in ICF), while fusion or fission materials (e.g., tritium or plutonium) can be produced using particle accelerators, or enriched using lasers. Similarly, in its direct application to military objectives, high-energy beam technology is important for directed-energy applications such as anti-ballistic missile defense and anti-satellite beam weaponry, while compact fission or fusion reactors are important for powering military ships, submarines, spacecrafts, and directed energy weapons.

Moreover, many ancillary technologies associated with thermonuclear fusion and high-energy beams are of great military importance: High-power radio-frequency generation for radars and electromagnetic warfare, superconductivity and cryogenics for outer-space military platforms, high magnetic field generation for magnetic compression and pure-fusion explosives, pulsed-power technology for flash x-ray radiography and electromagnetic guns, heatproof and heavy-irradiation-proof materials and electronic devices for conventional and nuclear explosives, micro- and nano-technology for mass-producing fourth-generation nuclear weapons, etc.

In the following sections several of these military applications will be described in some details.  They will generally come about as examples of implications (or ``spin-offs'') of various schemes which have a theoretical potential to become a source of thermonuclear energy if some intrinsic limitations are overcome.  Most of the time, these spin-offs will not be directly connected to nuclear weapons or proliferation \emph{per se}.  However, in all cases, it will be seen that the military and proliferation impacts are sufficiently important to be of serious concern, especially in non-nuclear-weapon States, because research on these ambiguous technologies can be legitimized by their theoretical potential as future sources of energy.

To conclude this section, and  before proceeding to a systematic analysis of fusion energy systems, it is important to remark that it is a legitimate right for all countries to pursue some exploratory research on all possible scientific developments that might have a revolutionary impact on military technology. In this respect, it is precisely in the area where fission, fusion and high-energy beam technologies overlap that the potential for new types of weapons is the greatest.  Moreover, if this set of now relatively standard technologies is extended to include more advanced atomic and nuclear processes (see Table 1),  we see that there is a relatively large number of physical processes available, which all have the potential to release large amount of energy on a short time scale, and which are therefore amenable to the design of new types of military explosives. This can be seen as a confirmation that atomic and nuclear physics are relatively new sciences, and an indication that many surprising discoveries are still possible, with many implications for new types of nuclear explosives.

\section{Principles of magnetic and inertial confinement fusion systems}
%=======================================================================

Since fusion reactions happen only at extremely high temperatures, the burning fusion fuel is a plasma, i.e., a collections of electrons and atomic nuclei from which the electrons have been stripped away.  Such a plasma cannot be contained by the wall of any vessel and must therefore be confined, either by gravitation as is the case in the sun\footnote{The thermonuclear reactions which ``power'' the sun are the ``carbon-cycle'' and ``proton-set'' fusion reactions. They are extremely slow and have nothing to do with the ``deuterium-tritium'' based fusion reactions used in nuclear weapons and in all practically feasible fusion reactors.}, or by electromagnetic fields if the reaction has to be continuous, or else by its own inertia if the reaction is pulsed (as is the case in nuclear weapons). 

This defines the two main categories of \emph{magnetic} confinement fusion (MCF) and  \emph{inertial} confinement fusion (ICF) systems. To these two generic categories one has to add many \emph{alternative} fusion systems (AFS) which use  the basic features of MCF and ICF in various configurations and combinations,\footnote{Hence the sometimes used term \emph{magnetic-inertial} confinement (MIC) fusion.} and in association with various technologies such as  pulsed electromagnetic power, energy cumulation, muon catalysis, etc.  Even though they are often called ``innovative,'' ``emerging,'' or ``new,''  most of these alternative or unconventional fusions systems have been extensively studied in military laboratories starting in the 1950-1960s, just like MCF and ICF systems.

Even though the details of \emph{magnetic confinement fusion} (MCF) systems are very complicated, the general concept is sufficiently simple and intuitive that we do not need to go into these details:  It is enough for our purpose to say that we have a magnetically confined high-temperature plasma steadily burning in a vessel which is heated by the energy released by the fusion reactions, and coupled to a turbine to make electricity.  While different kind of thermonuclear fuels could be considered in theory, its is likely that MCF is only feasible using deuterium-tritium as the fuel, so that the main thermonuclear reaction is $D + T  \longrightarrow  {^4\!H\!e} + n + (2.8 \times 10^{-12} {\rm J})$~.

The concept of \emph{inertial confinement fusion} (ICF) is that a sequence of tiny fuel pellets containing deuterium and tritium are projected towards the center of a reaction chamber where high-power laser or particle beam pulses strike each pellet, compressing and heating its fuel, and releasing thermonuclear energy by the reaction:  $D + T  \longrightarrow  {^4\!H\!e} + n + (2.8 \times 10^{-12} {\rm J})$~.  This nuclear energy is converted into thermal energy in an absorbing blanket lining the inner surface of the reaction chamber, and coupled to a turbine to make electricity.

Since 1 g of $DT$ produces about 340 GJ of energy, a nominal 1 GW (electric) fusion power plant with a thermal efficient of 30\% would consume 10 mg of $DT$ per second.\footnote{This corresponds to a consumption of 0.5 kg of tritium per day.} If we assume that one pellet is detonated each second, the explosive yield of each pellet would be 3.4 GJ, i.e., equivalent to about 810 kg of TNT.

Figure 5  is a simplified diagram of an advanced indirect-drive ICF target of the kind that is extensively studied for future ICF reactors. Such a target consists of a hohlraum containing a  1--10 mg  $DT$ fuel pellet.  The concept of \emph{indirect drive} refers to the fact that in this type of target the driver energy is not directly deposited onto an outer layer of the fuel but is first converted into thermal x-rays confined in a hohlraum. In the U.S., this concept was declassified in 1979 at the same time as the Teller-Ulam principle (Fig.3), using a wording that is almost identical: ``In some ICF targets, radiation from the conversion of the focused energy (e.g., laser or particle beam) can be contained and used to compress and ignite a physically separate component containing thermonuclear fuel'' \cite[p.103]{USDOE94}. 

It is therefore not surprising that  Fig.3 and Fig.5 are very similar, except for the technique used to generate the soft x-rays filling the hohlraum. In laser driven ICF, the hohlraum is generally a cylinder with openings at both ends to allow the laser beams to heat the inner surface of the hohlraum, causing emission of x-rays. In heavy-ion driven ICF, the heavy-ions are stopped in converters (i.e., small pieces of high-Z materials placed within the hohlraum) which are strongly heated. With other drivers, e.g., light-ion beams or antiprotons, the details would be different, but the result the same: strong heating of the radiation case or of the converters leading to  x-ray emission into the hohlraum. Hence, any type of indirect drive ICF system will enable the simulation of H-bomb physics in the laboratory.

A major problem, as for hydrogen bombs, is to succeed in compressing the pellet to a very high density and \emph{igniting} the fuel.  This is why the largest ICF system under construction in the United States is called the \emph{National Ignition Facility} because its main objective (which is same as those of its French and Japanese competitors) is to achieve ignition. (See Table 2 for a list of the most important laser ICF facilities in the world.)

The difficulty of this task is enormous, and it would be advantageous to find a technique similar to Teller's ``sparkplug'' concept which considerably simplified the design of H-bombs.  Several such techniques are under investigation, and one of the most promising is based on the invention of the ``superlaser,'' which enabled a factor of one million increase of the power of table-top lasers.

``Superlasers'' are ultra-short ultra-intense pulsed lasers with pulselengths in the range of $10^{-15}$ to $10^{-12}$ s, i.e., femtoseconds to picoseconds,\footnote{A femtosecond, i.e., 1 fs = $10^{-15}$ s is on the order of the time taken by an electron to circle an atom.  This gives the order of magnitude of the minimum pulselength of an optical laser pulse.} and beam intensities on the order of $10^{20}$ W/cm$^2$, i.e., sufficient to induce strong relativistic, multi-photon, nonlinear, and direct nuclear effects \cite{PERRY94, JOSHI95, MOURO98, TAJIM02}.\footnote{They are called \emph{super}lasers in this review because their interactions with matter are qualitatively very different from those of ordinary lasers.}  Their invention ended a twenty years long period over which laser intensity plateaued at a maximum of about $10^{14}$ W/cm$^2$.

The potential military applications of superlasers are so impressive that their principles have been implemented on existing large laser systems built for inertial confinement fusion and weapons simulation, pushing their peak power by three orders of magnitude from 1 TW to 1 PW which is why superlasers are more commonly called ``petawatt lasers.''  As can be seen in Table 3, there are now superlasers in operation or under construction in all major industrial states.

Superlasers enable a two step approach to ICF  similar to the ``sparkplug'' ignition of a cold compressed fuel in H-bombs \cite{TABAK94, PUKHO96}.  The proposed ``fast ignition'' scheme is as follows:  first, a capsule is imploded as in the conventional approach to inertial fusion to assemble a high-density fuel configuration; second, a hole is bored by a superlaser through the capsule corona composed of ablated material; finally, the fuel is ignited by fast electrons, produced in the superlaser-plasma interactions, which then propagate to the center of the pellet. This new scheme enables a factor of 10--100 reduction in total driver energy; it also drastically reduces the difficulty of the implosion, and thereby allows lower quality target fabrication, and less stringent beam quality and symmetry requirements from the implosion driver \cite[p.1626]{TABAK94}.

The whole subject of superlaser research and development is presently a domain of very intense activity.  New institutes and specialized laboratories  have been created in several countries. For example, the \emph{Center for Ultrafast Optical Science} at the University of Michigan, the \emph{Max-Born-Institut für Nichtlineare Optik und Kurzzeitspektroskopie} (MBI) in Berlin, the \emph{Centre Etude Lasers Intenses et Applications} (CELIA)in Bordeaux, or the \emph{Advanced Photon Research Center} (APRC) of the Japan Atomic Research Institute (JAERI) Kensai establishment.  As shown in Table 3, all the most advanced industrialized countries have now superlasers with powers of at least 10 TW, and 100-1000 TW superlasers in operation or under construction.

At present, with the closing and dismantling of the large American and French lasers to allow for the construction of NIF and LMJ, the most powerful lasers and superlasers in operation are in Japan and in England.  Japan is in fact the current world leader, and was able to announce on March 28, 2003, that  APRC ``succeeded in the development of a compact laser system capable of generating the world's highest laser output power of 850 trillion watts (0.85 petawatts)'' \cite{YAMAK03}.  Contrary to previous ``proof of principle'' petawatt-class superlasers which were quite large, this compact superlaser is based on solid-state Ti:Sapphire amplifier technology.

Germany, however, is proceeding very fast and trying to catch up with the rest of the world \cite{GSPON02}. This is done in collaboration with the Lawrence Livermore National Laboratory, with which Germany has passed ``agreements'' which were previously only available to England and France, two nuclear-weapon States \cite{GSPON02}.  In the the foreword of the \emph{GSI Scientific report 2002}, one reads :  "The highly appreciated delivery of components from the NOVA laser from Livermore gave a push to the PHELIX laser project (...) petawatt operation for end of 2004/beginning of 2005" \cite[p.ii]{GSI2003}.

Superlasers are an example of a breakthrough that is the result of pure technological innovation.  It was known since many years that one day a way would be found to go from the $10^{14}$ W/cm$^2$ standard laser intensity to the  $10^{20}$ W/cm$^2$ range because there is no \emph{fundamental} obstacle until the laser intensity limit of  $10^{24}$ W/cm$^2$ is approached.\footnote{In contrast, technologies such as thermonuclear fusion for electricity generation are \emph{already} at their scientific limit. No technological breakthrough is to be expected for them. This is why, assuming that all the details are worked out, whatever confinement scheme is adopted, it is already possible to compare these systems with other sources of electricity.}

However, there are other technologies (some of which listed in Table 1) which have also the potential of facilitating the ignition of an ICF pellets, and even to be sufficiently compact to be able to turn an ICF pellet into a fourth generation nuclear weapon.  One such technology is antimatter \cite{GSPON87a, PERKI97, PERKI03}, and several other possibilities (e.g., nuclear isomers \cite{WEISS93}) are discussed in reference \cite{GSPON97a}.

\section{Common proliferation implications of all fusion energy systems}
%=======================================================================

There are seven main common sources of proliferation associated with all types of fusion energy systems: neutron abundance, tritium abundance, new technological routes to uranium enrichment, latent thermonuclear weapon proliferation, induced nuclear weapon proliferation, and links to emerging military technologies.

\subsection{Neutron abundance: Fusion-fission hybrids and plutonium breeders}

While fission produces about 200 MeV of energy and between two and three neutrons per split nuclei, whereas fusion of $DT$ generates only 17.6 MeV of energy and a single neutron per fused nuclei, the fission reaction is called ``energy rich -- neutron poor'' and the fusion reaction ``neutron rich -- energy poor''.  Therefore, fission is most appropriate for producing energy, and fusion for producing neutrons.\footnote{This is why the most powerful nuclear weapons, i.e., those with the highest yield-to-weight ratio, use a ``third stage'' in which the fusion neutrons from the second stage are used to fission a uranium or plutonium blanket surrounding it.}

The consequence of this fundamental fact is that the most effective technique for producing nuclear \emph{energy} is to surround the reaction chamber of a fusion reactor by a blanket in which the fusion neutrons produce additional neutrons and breed plutonium, which is then burnt in ordinary fission reactors.  This is the concept of the ``fusion-fission hybrid'' \cite{HARMS80}, which takes advantage of the factor $200/17.6 \approx 10$ which favors fusion relative to fission as a supply of neutrons, and fission relative to fusion as a source of energy.\footnote{This factor of ten assumes that in a critical reactor one neutron is needed to maintain the chain reaction, and that the fate of the another fission neutrons is similar to those from fusion once the they are multiplied in a blanket.}

In the 1980s the main proliferation problem of these hybrids was felt to be the consequences of their exceptional potential as plutonium breeders \cite{GSPON83}.  This was because for a nominal power of 3000 MW (thermal) the best heavy-water fission reactor would breed ``only'' 500 kg/Pu per year, while a hypothetical fusion-fission hybrid of the same power would breed more than 5'000 kg/Pu per year.

Today, because of the existing surpluses of (both military and civilian) plutonium and enriched uranium, the emphasis is different:  a fusion hybrid is about ten times more effective than a fission reactor to breed plutonium or tritium.  In particular, for a given material output, its production of heat, nuclear waste, and traceable fission products such as $^{85}\!K\!r$ is at least ten times less than with a fission reactor. This makes such a facility easier to hide from inspection.\footnote{The only other systems which shows similar characteristics are the accelerator-breeders \cite{GSPON83}. These hybrids are now promoted for the efficient production of tritium for military purposes (see end of Sec.2.5.3).}  Moreover, the sub-critical blanket of a hybrid reactor is more appropriate for plutonium production than a critical fission reactor.  With an appropriate design, continuous extraction of plutonium is easy. Even for long neutron irradiation times in the blanket, the quality of the plutonium produced in a fast hybrid blanket is better than that produced in a critical reactor \cite{LEONA76}. A rather small facility, e.g., 10 MW(th) of fusion power, producing plutonium and tritium in a mixed uranium/lithium blanket, could be sufficient to maintain a minimal deterrent based on a few dozen high quality boosted fission bombs.

Of course, one may object that any fusion facility would be big, expensive, and very sophisticated.  That is not necessarily the case.  As will be seen in the section on alternate fusion systems, some of them might be quite small and rudimentary.  Moreover, if an easy route to ICF is found, small scale ICF might become feasible. 

One may also object that the $DD$ fusion reaction, and even more so the ``aneutronic'' fusion reactions such as ${^3\!H\!e}D$ and $^{11}Bp$, will generate much less neutrons. While this is true, it is also known since a long time, that
``the $DT$ reaction is the only one which can be used exothermically by the magnetic confinement method of the tokamak, as the other interesting reactions produce too high a loss of energy by cyclotron radiation'' \cite[p.331]{HORA91}.  The same is true for ICF: It is unlikely that any fuel other than $DT$ will enable the construction of economically attractive fusion power plants. In conclusion, the $DT$ fuel will always be the most attractive, and it will be a serious proliferation threat that a hypothetical future ``aneutronic-fusion'' reactor will switch to the $DT$ fuel cycle to produce neutrons for various illegitimate purposes.

\subsection{Tritium abundance: Boosted-fission and pure-fusion\\ nuclear weapons}

Whatever confinement-method used, a nominal 1~GW (electric) fusion power plant with a thermal efficient of 30\% would consume 10 mg of $DT$ per second. If we assume a tritium burn efficiency of 30\% (either in slow multi-second pulses as in MCF, or in fast pellet-explosions as in ICF), the amount of tritium to be processed and fed into the reaction chamber every day is 1.7~kg.  This tritium has to be regenerated and the idea is to breed this tritium in a lithium-blanket surrounding the reaction chamber.\footnote{The tritium breeding reaction is $n +  {^6\!Li}  \longrightarrow  T  + {^4\!H\!e}$.}  Assuming that at least one tritium atom is bred for every neutron incident on the blanket, and that about one tenth of the blanket is reprocessed every day to extract the tritium, a minimum of 5 kg of tritium will be constantly held in the blanket.\footnote{This is ten times the daily consumption of 0.5 kg of tritium.}  With a ten-day supply of tritium bottled to allow for reasonable down times, the total inventory is about $17+5 = 22$ kg.

These $\approx 20$ kilograms of tritium in every nominal 1 GW fusion power plant have to be compared to the  amount  of tritium in  a  military inventory, which in the case of the United States, is currently of about 100 kg. Since tritium has a half-life of 12.3 years, about 5 kg of tritium has to be produced every year to maintain this inventory.  Hence, from a vertical proliferation point of view, a single 1~GW fusion plant (which can easily breed more than the 0.5 kg of tritium needed every day for its own consumption) is more than sufficient to satisfy the tritium needs of a large nuclear arsenal, and even to envisage supplying a pure-fusion weapons arsenal which may require a significantly larger tritium inventory.

From a horizontal proliferation point of view, there are two major problems:

\begin{itemize}

\item Whereas to make one crude fission weapon about 3--6 kg of plutonium have to be diverted from a fission power-plant (i.e., about 1--2 \% of its fissile material inventory), to make a tritium-booster for one fission weapon only 2 g of tritium have to be diverted from a fusion power-plant (i.e., about 0.01 \% of its tritium inventory). Considering the existing difficulties with plutonium accounting, such a level of tritium control seems impossible to achieve in practice. 

\item Future nuclear arsenals will most probably be based on some kind of pure-fusion weapons. Such \emph{fourth generation nuclear weapons} will require only small amounts of tritium per warhead.  Assuming a relatively poor tritium burn-efficiency of 20 \% in these weapons, every  milligram of tritium would be worth the equivalent of about 100 kilograms of TNT.\footnote{We will come back later on the military value of highly compact weapons with yields in the hundreds to a few thousands of kg TNT.} Hence, assuming a modest breeding ratio of 1.1, a single 1 GW fusion plant would be able to supply 0.05 kg of tritium daily, enough to fabricate every day 5'000 warheads with a yield equivalent to 1 ton TNT each.

\end{itemize}

\subsection{Dedicated nuclear weapons materials production facilities}

Priorities in the present fusion programs worldwide emphasize the development of
MCF systems for electricity production and ICF systems for nuclear weapons simulation.  However, there have been publications (e.g., \cite{GSPON83, LOKKE86, MACIL96}), and certainly a number of classified studies, suggesting that such systems could also be used to breed large quantities of nuclear materials such as tritium and plutonium for military purposes.

In the nuclear-weapon States, the arguments put forward in favor of the construction of such facilities are (1) the need to replace aging facilities\footnote{In the U.S.A, for example, tritium was produced only at the Savannah River complex. The reactors at  Savannah River are now over 50 years old and will most probably never be restarted.} and (2) the increased importance of tritium relative to  fissile materials in modern nuclear weapons.\footnote{``The nation will have a continued need for an assured supply of tritium as long as nuclear weapons exits. [...] Moreover, certain changes in the makeup of the stockpile, underway or being considered, could actually increase the need for tritium even if the number of weapons declines.  Steps toward `modernization' of weapons, such as `safer' high explosives, or new missions calling for tailored effects, may lead to weapons with substantially higher than typical amounts of tritium.  Long-range nuclear materials planning requires assuring plutonium production as well'' \cite[p.2]{LOKKE86}.}

As was emphasized in the two previous subsections,  a fusion reactor can breed up to ten times as much tritium or plutonium than a fission reactor of the same power. Since disposal of heat is a major cost driver in any type of nuclear reactor, a major advantage of fusion-based systems is that they will produce much less heat for a given nuclear material output. This advantage becomes particularly important if a fusion machine is specifically designed to maximize neutron production for nuclear material breeding instead of heat production for electricity generation. This kind of optimization requires the development of a neutron multiplying blanket that poses fewers technical issues than a blanket for electric power generation.\footnote{This is also the case for the design of the blanket of an experimental fusion reactor such as ITER:  In a research facility heat is an unwanted byproduct of fusion complicating the design of the reactor and interfering with experiments such as the study of tritium breeding or of the effects of neutrons on structural materials.}

One approach, which minimizes heat production in the blanket, is to multiply neutrons by the $(n,2n)$ reaction in beryllium embedded in the blanket.\footnote{This concept is included in most ITER blanket designs, for instance the Japanese mass-production technology for ceramic tritium breeder pebbles under development at JAERI \cite{KAWAM00}.}  In this way, the blanket can be designed to breed 1.6 useful neutrons per fusion reaction. These neutrons can breed 1.6 tritons, and hence, an excess of 0.6 triton per fusion reaction for a total energy release of about 25 MeV (see Table 5). As is shown in this table, fission can also yield a net 0.6 triton per fission reaction, but with an energy release of 200 MeV.  Therefore, fusion produces eight times less heat than fission relatively to breeding capacity.

In small MCF tritium breeders this large cost advantage may be offset  by the cost of plasma technology (magnets, vacuum, etc.) required to confine the plasma. However, since plasma technology costs do not scale directly with fusion power, the capital cost of a magnetic confinement fusion breeder is relatively insensitive to thermal power and breeding capacity over a wide range \cite[Sec.2]{LOKKE86}. This leads proponents of this technology to conclude that ``because the fusion driver for a nuclear materials production facility would be smaller than a fusion power reactor, the present status of magnetic fusion research is, relatively speaking, more advanced for the purpose of breeding tritium than for energy applications  \cite[p.6]{LOKKE86}.''\footnote{This means that the blanket design for an experimental reactor such as ITER is better suited for breeding tritium than for producing energy.}

Similar arguments are used by proponents of inertial confinement fusion who stress that ``defense money will be unavailable [for a future ICF machine] unless it has a specific defense function such as tritium production or the destruction of plutonium'' \cite{MACIL96}.

In fact, while eventual future fusion reactors will pose a serious proliferation problem because of the possibility to divert tritium towards military ends, the declared nuclear powers will always prefer to rely on dedicated facilities to satisfy their military needs.  In this case, the main competition to fusion for supplying tritium in the future comes from accelerator based systems, which in addition to many advantages similar to those of fusion, have the interest to be a feasible extrapolation of existing proven technology.  In the event that such accelerator breeders are built in the near term, e.g., in the U.S.\ \cite{WEISM95,LAWRE98} or France \cite{GSPON97b,LAGNI98}, or possibly Japan \cite{MIZUM98, OYAMA03} or South Korea \cite{CHOI-03}, fusion will remain an attractive option for a more distant future.

\subsection{New or renewed enrichment technologies}

Natural uranium is a quite widely available commodity.  However, to make a fission-explosive, the difficulty is that natural uranium must either be burnt (e.g., in a heavy water reactor) to produce plutonium, or enriched to separate the fissile $^{235}U$ fraction from the more abundant non-fissile $^{238}U$ component.  Many technologies are available for uranium enrichment and some of the most efficient ones make use of lasers of various kinds.

While the large high-energy lasers used as drivers for ICF are not directly applicable to laser-enrichment, the underlying technology is the same.  Moreover, superlasers may open new  proliferation paths to laser-enrichment because they can be used to pump isomeric nuclear states for gamma-ray lasers, energy storage, and new military explosives \cite{WEISS93}.  When used to excite the first isomeric state of $^{235}U$, superlasers may lead to a very effective and compact technique for 100\% uranium enrichment in one stage \cite{OKAM77}.

Concerning other enrichment technologies, the ones that are based on separation in an electromagnetic field, i.e., the ``calutron'' method \cite{GSPON95} and the ``plasma separation'' process, are the ones that will most directly benefit from the development of high-current ion beams and superconducting magnets technologies for ICF and MCF \cite{GSPON83, GSPON95, HILAL88}.

\subsection{Latent thermonuclear proliferation}

Thermonuclear fusion is being promoted by the nuclear-weapon States as a potential source of energy and therefore a legitimate subject of study in non-nuclear-weapon States. This  leads to a growing situation of latent thermonuclear proliferation that will be examined in section 2.9.

\subsection{Induced nuclear proliferation}

Since thermonuclear fusion research and development creates the scientific and technical basis for actual or latent thermonuclear weapons proliferation, in either nuclear or non-nuclear-weapon States, it is legitimate for the less advanced countries to be concerned by this situation.  This may induce them to acquire the knowledge and skills for making first-generation nuclear weapons, which in turn become a justifications for the nuclear-weapon States to produce more advanced and usable nuclear weapons to effectively deter or counter this type of proliferation, and so on...

\subsection{Links to emerging military technologies}

As was already explained in section 2.3, all technologies relevant to thermonuclear energy have important military applications in nuclear and non-nuclear weapons.  Lasers, accelerators, superconductive devices, etc., are key components of new kinds of non-nuclear weapons: Laser and particle beam-weapons, electromagnetic guns, etc.

A clear example of the ambivalence of thermonuclear research with high-energy lasers is the project to use the technology developed for the U.S.\ National Ignition Facility (NIF) to build an Earth-based laser to clear near-Earth space debris \cite[p.15]{PHIPP96}: Such a device would also be an effective anti-satellite weapon.  Several other example will be cited in the following sections.

\section{Specific proliferation implications of inertial confine\-ment fusion (ICF)}
%========================================================================

The most direct proliferation contributions of ICF are obviously in the domains of thermonuclear weapons's physics and effects.  These implications have been expounded in a number of publications (see, e.g., \cite{GSPON83, GSPON97a} and references therein). Since the nuclear-weapon States are not shy at recognizing that ICF is essential to their nuclear deterrent, it is interesting to compare the headlines of two formal assessments of the relevance of ICF to the nuclear weapons program of the United States:

First, in 1977, fourteen years after the signing of the Partial Test Ban Treaty (PTBT), the main nuclear weapons's applications of ICF were presented as follows \cite[Chart 6]{STICK77}:

{\bf 

\begin{itemize}

\item Provides support for underground tests

\item Potential for large-scale weapon effects simulation

\item Allows modeling of atmospheric nuclear explosions

\item Provides unique capability for modeling of nuclear weapons physics including:

  \begin{enumerate}

  \item Normalization of design codes

  \item Studies of implosion physics

  \item Measurements of critical materials properties

  \end{enumerate}

\end{itemize}

}

Second,  in 1997, one year after the conclusion of the Comprehensive Test ban Treaty (CTBT), in a review of the ICF program, the section about the ``relevance of the National Ignition Facility (NIF) to Science Based Stockpile Stewardship (SBSS)'' is highlighting the following items \cite{KOONI97}:

{\bf

\begin{itemize}

\item People

\item Certification of the weapons stewards

\item Code validation and materials properties

\item Ignition:

  \begin{enumerate}

  \item The study of burn in the presence of mix

  \item The diagnosis of ICF plasma

  \item High-energy-density phenomena

   \item Proof-of-principle for inertial fusion energy (IFE)

  \end{enumerate}

\end{itemize}

}

Therefore, in both evaluations, the main emphasis is on showing how ICF is compensating for the perceived losses due to a major arms-control agreement, and the second on underlining the potential benefits of ICF for the future of nuclear weapons development.

In 1977, ICF was at its beginning. It was important to show that it would be a powerful substitute for \emph{atmospheric} tests and an effective simulator for weapons {effects} studies.  This substitution has progressively been done over the past twenty years (with the help of ICF and a number of pulsed-power technologies) and will be complete when large ICF facilities such as NIF will yield the very intense bursts of neutrons that only underground nuclear explosions could provide.\footnote{In the meantime, the most intense neutron simulators have been accelerator-based systems such as the LANSCE at Los Alamos, where the weapons neutron research (WNR) facility uses a spallation neutron sources driven by a storage ring to reach instantaneous fluxes of $10^{22}$ n/cm$^2$/s \cite{GSPON83}.}

In 1997, SBSS is at its beginning. The age of building and testing multi-kiloton to multi-megaton hydrogen bombs is gone, and a new generation of people is needed:  People to take care of the existing arsenal without \emph{underground} tests, and people to design and possibly build the next generation of nuclear weapons using computer simulations and laboratory tests. Before coming to the implications of ICF for this fourth generation of nuclear weapons, let us review the more trivial military applications of ICF:

\subsection{Nuclear weapon-effects research}

ICF systems enable both nuclear and non-nuclear effects to be studied.  The latter consists of the effects of low and high altitude single and multiburst detonation in the atmosphere.  Such studies enable (a) prediction of the effects of subsequent bursts in a multiburst environment; (b) evaluation of the spatial extent and duration of satellite communication interference; and (c) evaluation of radar shielding effects which hinder detection of secondary missions \cite{TOEPF78}.  Since 1964, because of the PTBT, such problems cannot be studied with real nuclear explosions in the atmosphere.

The total radiation field of a nuclear explosion is composed of x-ray, gamma-ray, neutron and electromagnetic pulse (EMP) components.  The intensity of each of these is strongly dependent upon the specific design and the yield of the weapon.  Also, the presence or absence of some of these radiations depend on the environment in which the nuclear detonation occurs.  For example, in an underground explosion some of the radiation (e.g., EMP) will be absent compared to an atmospheric or high altitude explosion \cite{TOEPF78}.

Until the conclusion of the CTBT, synergistic testing was done through underground explosions, but ICF provides now an alternative method  for carrying out such tests in the laboratory; an ICF exposure is expected to cost less than one percent of an underground experiment.\footnote{However, for countries such as India, because of the complexity and huge investment cost of large ICF facilities, underground tests are potentially less expensive than ICF simulations.}  Furthermore, experiments with an ICF facility are much more convenient and reproducible.  For example, meter-sized costly equipments such as reentry vehicles, missiles, satellites, can be exposed to neutron fluxes of  $10^{13}$ to $10^{14}$ n/cm$^2$/s, or 3 to 30 cal/cm$^2$ x-rays, without completely destroying them.  ICF systems can also be used for ``nuclear hardening,'' and to ``burn in'' ready-to-field equipments by exposing them to radiations and replacing the weakest components that may have failed.

\subsection{Nuclear weapons-physics research}

After the discovery of the Teller-Ulam principle, and some major improvements  during the 1960s, progress on thermonuclear weapons  slowed down dramatically. In fact, despite more  than  50  years  of  research  and  development, and  after almost two thousand test explosions,  the  scientific  understanding of the details of  the  secondary  system  is  still  incomplete.\footnote{``We do not completely understand the physical processes involved in the operation of a nuclear weapon''~\cite[p.24]{LIBBY94}. ``We do not understand nuclear weapon processes well enough to calculate precisely the transfer of energy within a weapon''~\cite[p.30]{LIBBY94}. For a more detailed development of the subject of this subsection see, in particular, \cite[Chap.3]{GSPON97a}.}  If the CTBT would not have been concluded, the continuation of full-scale testing would probably never have changed this situation, given the great number of complex phenomena that occur simultaneously within the fraction of a microsecond of the explosion of an H-bomb.

A major problem with full-scale testing is that the secondary of an actual bomb is buried deep inside the weapon, surrounded by a thick ablator and the radiation case. Therefore, most experimental data on the thermonuclear part of the explosion is indirect.  In comparison, an ICF pellet is an almost naked secondary, and many configurations can be tested at will, with much better diagnostic capabilities than with underground nuclear tests.  The promise of ICF is a complete description of thermonuclear weapons physics from first principles.

Laser and particle beams are capable of concentrating large amounts of energy onto small targets.  These targets may consist of non-nuclear materials, fissile materials, or fusion materials.  The very high pressures and shock strengths possible with the kind of beams necessary to drive ICF systems, enable hydrodynamic behavior and material equations of state to be studied in a parameter range comparable to that existing within exploding nuclear weapons (see Fig.6).  The large megajoule-scale ICF facilities currently under construction (e.g., NIF, LMJ, see Table 3) will be particularly well suited for this purpose \cite{CAMPB97}, as well as smaller facilities such as Helen in the United Kingdom \cite{ONION02}.

But proton beams from high-power generators, such as, for example, the \emph{Karlsruhe Light Ion Beam Facility (KALIF)} (see Table 4) in Germany enable similar measurement with power densities of up to 200 TW/g and energies densities of several MJ/g \cite{BAUMU96}. As can be seen on Fig.6, such energies densities, equivalent to $2 \times 10^{-4}$ \emph{kt}/kg, are only a factor of 100 less than NIF without ignition. Other pulsed-power beam generators, such as the Saturn electron accelerator \cite{OLSON97} in the U.S.A.,  are also providing important nuclear weapons data, even though they are primarily advocated as fusion energy research tools.\footnote{Laser facilities, however, provide a much ``cleaner'' working environment, especially with respect to electromagnetic interferences from the pulsed-power equipment, and much better energy deposition profiles because of the greater flexibility of time-shaping the laser pulses.}  This is also case of PBFA (or ``Z Machine'') at Sandia National Laboratory (see Table 4) which in 2003 was able to produce a significant amount of DD fusion neutrons that could previously only be generated with laser beam facilities \cite{SCHWA03}. 

Moreover, lower-energy kilojoule-scale laser facilities equipped with a superlaser for fast ignition --- such as the 60 beams 50 kilojoule Japanese laser FIREX-II (see Table 3)\footnote{FIREX stands for ``Fast Ignition Realization Experiment.  For a description and planning of the FIREX project, see the \emph{Annual progress reports} 2001 and 2002 of the Institute of Laser Engineering, Osaka University.} combined with a 4 beams 10 kilojoule petawatt superlaser --- are also expected to achieve ignition, and therefore to enable thermonuclear burn studies comparable to those feasible with NIF or LMJ (see Fig.6).

The complexity of ICF target experiments requires that they be analysed by simulating the experiment with two- and three-dimensional hydrocodes.  Thus verification and improvement of weapon design code is an intrinsic part of ICF experiments.  Since ICF research is done in non-nuclear-weapon States, very sophisticated computer codes have been developed and published by scientists in such States. For instance, the two-dimensional hydrocode MULTI2D \cite{RAMIS92} developed at the \emph{Max-Planck-Institut für Quantenoptik}, in Garching, Germany, is considered to be in several respects better and faster than LASNEX, the currently standard (and partially classified) U.S. two-dimensional hydrocode.  These codes allow, in particular, the simulation of the dynamics and stability of implosion (of either passive or nuclear materials) driven by x-rays, high-energy beams, or other types of drivers: Chemical high-explosives, magnetic fields, electromagnetic guns, etc.

Considerable scientific data necessary for the design of fusion systems is also crucial for thermonuclear weapons.  For example, the temperature- and pressure-dependent opacity functions for high atomic-number elements were classified until 1993 because this information is needed to make such weapons.  Techniques for measuring these opacities are improving because of the availability of high-energy lasers.  These can be used to measure opacities directly at laser frequencies, or indirectly by converting the laser radiation to x-radiation, and measuring opacities in the x-ray region which is the most relevant to nuclear weapons.

A last aspect of ICF which is of importance in weapons physics is that of rate-dependent processes.  An ICF system can easily expose a recoverable target to neutron and x-ray fluxes comparable to those of a full size nuclear explosion.

\subsection{Ignition: Fourth generation nuclear weapons and inertial fusion energy (IFE)}

``Ignition,'' the last item stressed in the 1997 assessment of ICF, refers to the fact that macroscopic processes like thermonuclear plasma \emph{ignition} are still not well understood. Special ICF targets which absorb the driver energy and convert it to x-rays enable H-bomb ignition physics to be studied directly.  Moreover, ignition techniques different from the Teller-Ulam concept can also be studied, or discovered, using ICF. ``Fast ignition'' using a superlaser, or the replacement of the ``sparkplug'' by a minute amount of antimatter \cite{PERKI97}, are examples of such new concepts.

The important conclusion is that, whatever the details, successful ignition of thermonuclear micro-explosions in the laboratory will open the way to two types of applications which will most certainly remain in the military domain, i.e., new types of nuclear weapons, and inertial fusion energy (IFE):

\begin{itemize}

\item \emph{Fourth generation nuclear weapons.}  Inertial confinement fusion is basically a continuous salvo of contained thermonuclear explosions.  In a nominal 1~GW fusion power plant, the yield of these explosions will be equivalent to about 100 or 1000 kg of TNT, assuming a rate of 10 or 1 detonations per second, or else to about 10 tons of TNT, assuming a rate of one detonation every 10 seconds. The military significance of these yields derives from the fact that the amount of conventional high-explosives (HE) carried by typical warheads or bombs is limited to a range of a few 100 kg to a few tons.\footnote{The Scud ballistic missile warhead contains roughly 200 kg of HE and the Patriot anti-missile warhead roughly 40 kg. A Tomahawk long-range cruise-missile carries a conventional or thermonuclear warhead weighting about 120 kg, and a typical big air-dropped bomb weighs between 500 and 2000 kg.}  Since an ICF pellet weighs only a fraction of a gram, ICF based military explosives would revolutionize warfare.  Combined with precision guidance, earth and concrete penetration, and other existing techniques, small and lightweight ICF based warheads would destroy virtually all possible targets, and render existing types of high-yield nuclear weapons obsolete.  The challenge, of course, is to replace the large laser- or particle-beam driver by some sufficiently miniaturized device.  This problem will not be discussed here.\footnote{For a development of this question see \cite[Chap.4]{GSPON97a}.}  However, it will be recalled that a single-use device is usually much more compact and simple than a multi-purpose re-usable experimental facility, and that very-high energy-density technologies such as antimatter and superlasers are ripe to meet the challenge \cite{GSPON97a}.\footnote{These considerations about radically new types of nuclear weapons should not minimize the potential of using ICF facilities for improving existing types of nuclear weapons \cite{PAINE97}.}  In particular, in a review of the use of antiprotons as a compact ICF driver \cite{PERKI97} it has been found that the number of antiprotons required for the fast ignition of ICF pellets is several orders of magnitude smaller than was estimated by the authors of the present review ten years earlier \cite{GSPON87a}.

\item \emph{Inertial fusion energy.} Success with ignition, and a sufficient reduction of scale of the driver, would provide a very attractive substitute for the numerous nuclear reactors used by the military.\footnote{There are in the world more military nuclear reactors at sea, under the sea, and in various military facilities, than nuclear power reactors producing electricity for civilian purposes.}  As for the possible \emph{civilian} use of IFE, the prospect is bleak. Considering the bad image that nuclear energy has in general, it is unlikely that IFE will be found acceptable by the public in democratic countries.  For one thing, development of IFE will come in parallel with fourth generation nuclear weapons which will use ICF pellets as their main explosive charge.  A daily load of ICF pellets for a medium or full scale fusion power plant will consist of thousands of pellets, each of them equivalent to one or several tons of high explosives.  If these pellets are not fabricated at the power plant, their shipment will have to be heavily guarded.  Moreover, because fusion pellets contain only weakly radioactive materials, e.g., tritium, diversion will be difficult to detect.

\end{itemize}

\subsection{People}

As Ted Taylor\footnote{To whom we dedicated our report on fourth generation nuclear weapons \cite{GSPON97a}.} likes to say: ``The most important things to make a bomb are dedicated people and a good library.'' This is made clear by the emphasis put on ``people'' and ``weapons stewards'' by the U.S. National Academy of Science in their 1997 assessment of the relevance of ICF to SBSS \cite{KOONI97}. The trouble is that this is also true for the non-nuclear-weapon States. World-wide thermonuclear energy research is creating a basis which allows the much more rapid development of quite advanced thermonuclear weapons than would otherwise be possible.

\section{Specific proliferation implications of magnetic confine\-ment fusion (MCF)}
%========================================================================

Much of what has been said about ICF in the previous section applies to MCF, with the obvious differences which derive from the fact that the operation of a MCF device is characterized by a much smaller plasma density (and a correspondly much longer burn time) than in an ICF device or a thermonuclear explosion.  For this reason, MCF physics and technology is less closely related to thermonuclear weapons than is the case for ICF.  However, there are several areas in which MCF is providing specific direct or indirect contributions to nuclear weapons proliferation.  Let us briefly mention two of them:

\subsection{Nuclear weapons effects}

Tokamak test reactors have been proposed for nuclear weapon-effects research \cite{JASSB82}.  The main advantage is the very large volume with uniform irradiation flux of $10^{13}$ n/cm$^2$/s offered.  However, because of the very long pulselength compared with ICF, tokamak test reactors are best suited for determining the permanent damage due to neutron (and x-ray) radiation to total-dose-dependent rather than dose-rate-dependent devices and subsystems.  In 1982, it was hoped that the Princeton tokamak fusion test reactor could be used to help define an optimized tokamak radiation effects facility that could have been implemented by the early 1990s in a suitable location \cite{JASSB82}.

\subsection{Military spin-offs of MCF technology}

In a compilation of several surveys, it was found that the most numerous technology transfers from magnetic fusion research to other areas of science and technology were in the domains of magnet technology, power supplies, materials technology, particle beams, power supplies and vacuum technology \cite{UMEZU90}.  These types of spin-offs are similar to those found for high-energy particle accelerator physics research, a domain which like MCF produces relatively few spin-offs of direct importance to industrial development.

  Superconductive magnet are of great importance for strategic military developments in outer space and ballistic missile defense, as well as for tactical developments such as electromagnetic-pulse generators and electromagnetic guns. The use of cryogenic and superconductive magnets in space is been investigated for pulsed-power generation, power conditioning and energy storage,  and was expected to play a major role in the Strategic Defense Initiative (SDI) program \cite{LEUNG87}.

Superconductive magnets are also of great interest for the ``plasma separation process'' \cite{HILAL88} which is potentially the most attractive technique for very-high throughput isotope separation  \cite{TRACY89, THIKO92}.  Large scale isotope enrichment is important  for the production of $^{235}U$ as well as for the enrichment of various types of medium weight nuclear species if ``isotopic tailoring'' becomes an important feature of new materials.  Such materials are expected to be necessary for making the first wall of MCF fusion vessels (in order to minimize their erosion and radioactivation due to the intense neutron bombardment from the burning plasma) as well for a number of military applications where similar or related properties are important.

\subsection{Examples of spin-off technologies expected from ITER}

To conclude this section, we quote \emph{in extenso} the examples given in section 5.13 devoted to the spin-off benefits of fusion technologies in the summary of the report of the \emph{Special Committee on the ITER Project of the  Japanese Atomic Energy Commission}.  This is not to imply that the examples given by the Committee are necessarily relevant to the proliferation of nuclear weapons, but an illustration that they are indeed mostly dual-purpose technologies of great military significance:

\begin{quote}
``\emph{Driving force of spin-off technologies}

Since fusion development requires gathering knowledge from a myriad of advanced technologies, it is now making significant progress as a seed of these technologies. The fusion device is based on diverse research fields and fashioned from advanced technologies, such as physics, mechanical engineering, electric and electronic engineering, materials engineering, thermodynamics, heat transfer flow and thermal engineering, nuclear engineering, cryogenic engineering, electromagnetic dynamics, chemical engineering, and control engineering and instrumentation. Therefore, the development of this compound technology not only advances individual fusion technology but also raises the potential capability of all science and technology by mutual stimulation between different fields of science. The resultant spin-off benefits are seen in commercial technologies, such as the semiconductor industry and the large, precision machine-tool industry. Fusion research also contributes to the development of advanced technology and basic science of other fields, such as physics, space science, materials science, medicine, communications, and environmental science. These applied sciences include accelerator technology, superconductor technology, diagnosing techniques, plasma application technology, heatproof and heavy-irradiation-proof materials technology, impurity removal techniques, and computer simulation techniques. 

\emph{Examples of spin-off technologies}

Examples of spin-off technologies include the development of large superconducting coils for ITER, which reduced the cost by 75\% of niobium/tin superconducting wire material necessary of the generation of the high-magnetic fields. This has allowed the high-magnetic field MRI used for medical diagnostics to become relatively commonplace. At the same time, the AC loss has been reduced by 80\% of that for conventional superconductors, even at the strong magnetic field of 13 tesla. This makes it feasible to increase the stored energy in a superconducting power storage system by a factor of 5--7 when compared with a system designed with conventional technology and operating at 5--6 tesla. In addition, vacuum pumps for high thermal efficiency refrigerating machines, which operate below 4 K, have been developed and have been adopted at the Fermi National Accelerator Laboratory in the US and CERN in Europe. This also demonstrates the enormous contribution of fusion research to the frontiers of science. The technology of producing large positive-ion-beam currents, originally developed for the heating of fusion plasmas, has already pervaded into the technologies for products used in daily life, such the semiconductors used in the home electric appliances. In addition, the large negative-ion-beam current technology developed for ITER is expected to give birth to completely new research fields, such as the creation of previously unknown materials. The negative-ion beam, which has monochromatic energy, is also suitable for manufacture of intricate semiconductor devices. This allows the realization of low-cost, mass-produced single crystal silicon thin films for solar cells. Furthermore, high-power radio-frequency sources used for plasma heating are already applied to the manufacture of high-performance ceramics. Potential applications of these sources extend from solving environmental problems to the radar used in outer space. The integration of component technology for the fusion reactor also advances the systematic development of technologies addressing integration, such as system engineering, control engineering, and safety engineering. Additionally, an exploratory investigation related to the processing of radioactive waste by utilizing a fusion reactor itself as an intense neutron source is also being carried out and seems promising'' \cite[p.274-275]{RSC-ITER}.
\end{quote}

\section{Proliferation implications ~of~ alternative fusion systems (AFS)}
%=======================================================================

Besides the two main contemporary contenders to fusion power, i.e., MCF and ICF, a number of more or less attractive alternative schemes have been proposed: pinch devices, magnetized fuels and magnetic compression devices, chemical explosive driven systems, plasma focus, impact fusion, etc.

Like MCF and ICF, these concepts have generally been first studied at nuclear weapons laboratories because it was important to carefully evaluate their military potential before letting the scientific community at large working on them.

A particularly interesting example from the point of view of horizontal thermonuclear weapons proliferation is the so-called \emph{plasma focus} which has been independently discovered in Russia in 1962 and the United States in 1964.\footnote{The dense plasma focus is also known under the name of ``Mather gun'', from the name of its American inventor, Joseph W. Mather, who recently passed away. In his obituary  \cite{AHLUW98}, the $DD$ neutron yield of the most powerful DPF device known to have been built is mistyped $10^{12}$ instead of $10^{18}$.} Essentially, this is a fast dynamic Z-pinch in which the stored magnetic energy is rapidly converted into plasma energy and then compressed by its own magnetic field. It consists of two cylindrical electrodes between which a powerful electric discharge is initiated with a capacitor bank in a $DD$ or $DT$ atmosphere, and is therefore the simplest high-flux fusion neutrons and x-rays generator that exists \cite[p.172]{GSPON83}.

In the United States, plasma focus development culminated in 1974 with a device called ``DPF 6 1/2'' which produced $\approx 10^{20}$ $DT$-fusion neutrons at a repetition rate of about four pulses per hour.  If such discharges could be generated at a higher rate and over long periods of time, a hybrid system could be built, producing tens of kilograms of plutonium (or hundreds of grams of tritium) per year \cite{FISCH78,GRIBO80}.  While this possibility seems to be relatively remote at present, and will almost certainly never be cost effective, it should not be excluded as a simple path to a nuclear weapon capability, similar in kind to the ``calutron'' enrichment route taken by Iraq in the 1980s.

Because of its simple construction, and its richness in plasma physics phenomena, dense plasma focus devices have been constructed in many advanced (e.g., \cite{SCHMI84}) and developing countries.  Looking at the recent literature, the latter include, for example, India \cite{ROUT991}, Egypt \cite{SHARK94}, Malaysia \cite{MOO1995}, etc.  The construction and the operation of these devices, as well as the analysis of the data provided by them, allow to train physicists specializing in experimental or theoretical plasma physics, and give the possibility to travel abroad to actively participate in international conferences. They may also provide some indigenous expertise in a number of technical aspects connected with the physics  of simple thermonuclear weapons, including deuterium and tritium handling technology.

\section{(De)classification and latent proliferation}
%====================================================

History of the past 50 years shows that most scientific and technical knowledge pertinent to nuclear or thermonuclear weapons and reactors has been kept secret as long as possible by all countries which had a lead in these areas. Similarly, history shows that many (and possibly all) other countries were interested by nuclear and thermonuclear energy \emph{primarily} because of their \emph{military} applications, and only secondarily because of their potential applications as sources of energy.

The process of declassification of information by the nuclear-weapon States is therefore guided by political considerations in which military and foreign policy reasons play the most important role. For example, as is shown by the analysis of  President Eisenhower ``Atoms for Peace'' initiative of 1953, and of the Western Europe reaction to it, the atoms-for-peace policy of the United States was a foreign policy concept --- neither an energy policy nor a trade or economic policy \cite{KOLLE94, KOLLE96}.  Its main rationale was to demonstrate `world leadership' in order to bind neutral or politically indifferent countries to the West. As a reaction to this sensational turn in Washington's nuclear attitude, Moscow began shortly thereafter to join the proliferation of basic nuclear know-how and equipment --- although only within its own hemisphere, for instance China and Eastern Europe.  Consequently, most countries with sufficient resources launched a nuclear power development program which gave them a politically correct justification to achieve a \emph{latent nuclear weapon capability}, which could become an actual capability if needed at some point.

In the area of thermonuclear fusion, the secrecy that had been imposed on its development was lifted in the mid-1950s when an outstanding Russian scientists, I.V. Kurchatov, presented in England results obtained in the Soviet Union with the magnetic pinch device. This led the United Kingdom to declassify part of its controlled thermonuclear fusion research with the publication, in 1957, of a series of six articles in the \emph{Proceedings of the Physical Society} comprising the fundamental paper of J.D. Lawson defining the ``Lawson criterion'' for break-even in thermonuclear fusion \cite{LAWSO57}.  This in turn helped to persuade the American authorities to declassify the whole magnetic confinement approach to fusion.  At the second Atoms for Peace Conference in 1958, a complete disclosure was made by the United States and was fully reciprocated by the Russians.

This sequence of events illustrates that the Soviets decided to use their leadership in MCF research as part of their answer to Eisenhower's atoms-for-peace initiative. They did that because they felt that while their ``tokamak'' device was far better than anything in the West, forcing declassification in that area would not reduce their advance in the field.

From then on, more and more information was declassified, mostly following the publication of a growing number of sensitive results in the scientific and technical literature by an increasing number of researchers from all over the world. While possibly many of these results have been published because of non-existent or ineffective classification policies in some countries, most of them have been deliberately published (or not classified) for various reasons.  In particular, many significant research results have been published by scientists simply in order for them or their country to get credit for being the first to have discovered something.  Similarly, as an element of their deterrence policy, threshold nuclear States (especially India and Israel) and latent nuclear powers\footnote{In the late 1970s and early 1980s Swiss scientists published some of the most detailed simulations of the dynamics of fission explosions that have ever been published. This was at a time when Switzerland was still ``defending'' its status of a threshold nuclear-weapon State \cite[p.70]{STUSS96}.  Later, in 1985, the German Ministry of defense approached the Swiss government to obtain copies of several technical reports on nuclear and thermonuclear weapons that had been produced by Swiss scientists \cite[p.81-82]{STUSS96}.} tend to display their level of understanding of the physics of nuclear weapons by publishing results that have not already been published by others.

In the last twenty years, however, the main push towards the declassification of thermonuclear weapons physics has come from the scientists of the two major non-nuclear-weapon industrial powers: Japan and Germany. Japan, in particular, has been so vigorously pursuing its very ambitious ICF program that it has found itself in the position of the world leader at several occasions.  In 1986, for example, a thermonuclear neutron yield of $10^{12}$ was achieved with the Gekko XII laser --- a record that the U.S.\ and French nuclear scientists took several years to surpass.  Today, Japan is still leading the world with a  pellet compression of 600 times its initial solid density, a record achieved in 1989--90.  Similarly, in 1998, Germany has been the first country in the world to publish unambiguous results showing that thermonuclear reactions can be induced by a table-top size superlaser \cite{PRETZ98}, something that military laboratories in nuclear-weapon States had most probably already done before, but not published for obvious reasons. Further progress in that direction was than made by Japan, which was first to report in 2002 a considerable neutron yield enhancement with fast heating a compressed ICF pellet by means of a petawatt superlaser beam \cite{NATUR02}.

On the theoretical side, where both Japan and Germany are publishing everything, important results have been disclosed starting around 1982, when Jürgen Meyer-ter-Vehn published his now famous energy gain model of fusion targets \cite{MEYER82}.  For instance, many detailed analytical and numerical simulations\footnote{Using very advanced two-dimensional hydrodynamic simulation codes, e.g., \cite{RAMIS92}.} of indirect-drive  ICF, which use methods that are directly applicable to x-ray driven thermonuclear weapons secondaries, have been published by Japanese and German scientists, years before the corresponding calculations had been declassified in the United States or France.

Consequently, when the United States Government decided in 1993 to declassify formerly secret material concerning the indirect-drive approach to ICF, only a small part of what was declassified was really new to the open scientific community.  On the other hand, many European researchers in the field expected that the \emph{U.S. Government declassification act of 1993} would increase the priority given to ICF by their authorities.  For example, in justifying its recommendation to increase the European Union funding of ICF, a European Science and Technology Assembly (ESTA) committee  appointed in 1995 referred to this \emph{declassification act} as if the mere fact of declassifying something would remove its military potential \cite{ROJO96}. Since this recommendation was not followed by the European Union (possibly because of the opposition of France and the United Kingdom) some scientists were led to qualify their governments as being ``irresponsible'' (see, e.g., \cite{BLUHM97}). 

The problem is that latent \emph{thermonuclear} proliferation, the fact that countries such as Germany and Japan are able to build thermonuclear weapons on short notice, and even to eventually equip themselves directly with fourth generation nuclear weapons, bypassing the acquisition of previous generations of nuclear weapons, is not a trivial situation at all. In particular, it is not the result of any recent short-term political decision. It is neither something that can be changed by trivial political action.

Latent thermonuclear proliferation is the result of political decision made in the 1940s and 1950s which gave to (thermo)-nuclear science and (thermo)-nuclear energy the privileged status they still have today.  Because advanced thermonuclear research is now part of the normal scientific activity of all large industrial countries, their emergence as latent thermonuclear powers was unavoidable. In other words, latent thermonuclear proliferation and the possibility of fourth generation nuclear weapons in non-nuclear-weapon States is the result of a \emph{political and technological dynamics} which is today to a large extent out of the control of both the scientific and political communities.  In particular, this dynamics has \emph{institutionalized} thermonuclear energy research in such a way that it is very difficult to stop it, even though it is more and more overlapping with the most advanced thermonuclear weapons related research done in the nuclear-weapon States.

An aggravating circumstance is the increasing reliance on virtual capabilities by both nuclear-weapon and non-nuclear-weapon States.  This is the direct consequence of the unwillingness of the nuclear-weapon States to give up their nuclear weapons, and of their decision to replace actual tests by ``simulation'' and ``science based stockpile stewardship'' programs. Instead of having unambiguously kept or rejected the nuclear option, the nuclear-weapons-states are therefore lead to foster ``visible technical deterrence.''  In the words of the director of the nuclear weapons technology program at the Los Alamos National Laboratory:

\begin{quote}

``As I see it, at the end of the Cold War the nuclear weapons laboratories took a new mission.  Not only must we maintain the stockpile, but we must do so in a manner visible to the world, a manner that demonstrates our technical competence in scientific and engineering fields that are obviously related to nuclear weapons'' \cite[p.2]{YOUNG97}.

\end{quote}

This is precisely what Japan and Germany (and to a lesser extent some smaller countries such as South Korea) are also able to do without having to spend any resource on the maintenance of an expensive stockpile of high-yield nuclear weapons, which is why the continuing expansion of their thermonuclear research programs, and specific issues such as the possible siting of ITER in Japan, have such important political and strategic implications.

\section{Conclusions}
%====================

In this review, we attempted a systematical analysis of the actual and latent nuclear weapons proliferation implications of thermonuclear energy systems.

The first conclusion is that the main findings confirm those of earlier reviews of the subject (e.g., \cite{GSPON83}).  For instance, the development of thermonuclear energy systems (especially those based on ICF) enhances the knowledge of thermonuclear weapon physics, provides  an impetus for the development of a number of technologies which have mostly military applications, and leads to new methods for producing large quantities of military useful nuclear fuels such as tritium.

The second conclusion is that the successful development of any kind of thermonuclear fusion power plant would pose very serious nonproliferation problems because very large amounts of tritium (which are on the order of kilograms per day)  would be produced in these plants.  Moreover, if the plants would be based on inertial confinement fusion, operation would require the detonation of many thousands of ``micro hydrogen bombs'' every day, each of them with a yield measured in tons equivalent of high explosives.  Similarly to tritium, these pellets would have to be safeguarded because they could possibly be  used as very compact military explosives which would revolutionize conventional and nuclear warfare.

The third conclusion is that the widespread availability of supercomputers, and numerous scientific/technical advances such as the discovery of the superlaser, tend to decrease the scientific and technological gap between the nuclear-weapon and non-nuclear-weapon States. The result is that thermonuclear energy research enhances the competence of the most advanced non-nuclear-weapon States to the point where their status of latent thermonuclear powers becomes more and more equivalent to one of actual thermonuclear powers.

\chapter{Bibliography}
%=====================

\chapter{Appendix: ITER siting politics}
%============================================

\vspace{-1cm} \noindent Excerpt from a commentary on version 2 of this report by

\begin{center}
{\bf Clifford E. Singer}\\
\emph{Program in Arms Control, Disarmament, and International Security}\\
University of Illinois at Urbana-Champaign\\
%24 January, 2004
\end{center}

\noindent ITER siting is a matter I have been interested in for over 20 years.  The overall impact on siting in Japan depends heavily on two questions:
\begin{enumerate}
\item If ITER is not sited at Rokkasho, can it be built at all?
\item If ITER is sited at Rokkasho, will this be in addition to or as a 
substitute for operation of the spent nuclear fuel reprocessing plant there?
\end{enumerate}

\section{If ITER is not sited at Rokkasho, can it be built at all?}

Even though the decision to build ITER has nominally been made, it 
would not be surprising to see the needed financial commitments 
partly or completely fail to materialize if it is sited in France. 
Compared to Rokkasho, choice of a French site could 
substantially reduce Japanese enthusiasm for the project and 
perhaps also that of the United States as well. A Canadian site is 
also problematic due to potential problems with Japanese and European 
funding, even though U.S. scientists would find a Canadian site 
somewhat more convenient.

 From an energy point of view a near-term decision to build ITER is 
unlikely to be a cost-effective use of R\&D investment with any 
plausible non-zero discount rate and likely future scenarios for 
evolution of the cost and use of alternative energy technologies. 
However, from a basic science point of view tokamaks provide a 
uniquely interesting research tool. Tokamaks are particularly useful 
for studying transport in turbulent fluids, because turbulent 
transport is potentially calculable over all of the relevant length 
scales --- unlike for non-ionized gases in macroscopic systems. However, 
it is more cost effective to research this in much lower-cost 
tritium-free plasmas with careful attention to axisymmetry and 
careful diagnostics of deviations from it due to neutral gas 
recycling. Thus, if a Japanese site is chosen and is essential to 
actual construction of ITER, the effect on magnetically confined 
plasma research may be deleterious unless there is such a large 
overall funding increase that this precipitates a stronger research 
program using other tokamaks not burdened with radioactivity 
management that results from extensive high-power tritium operations.

In addition, a decision to abandon large-scale use of tritium in 
magnetic confinement research could gradually lead to a decoupling of 
the largely implicit connection to military relevance in the minds of 
policy makers. On the other hand, if ITER is to be built anyway 
regardless of whether a Japanese site is chosen, then the impact on 
plasma physics research will be about the same, and the relevant 
issues on where to site will be more as described in the paper to 
which these comments are appended.

\section{If ITER is sited at Rokkasho, will this be in addition to or as a substitute for operation of the spent nuclear fuel reprocessing plant there?}

It is useful to keep in mind that the cost of operating the Rokkasho 
reprocessing plant has been estimated at four billions dollars per 
year. According to Paul Leventhal who quotes this estimate in the 3-5 
December 2003 conference paper Verification of Nuclear Disarmament 
on the Korean Peninsula, ``Japan  could save at least half the cost of 
its plutonium-based program if it were to forego separation of 
plutonium and operation of plutonium-fueled reactors and instead 
developed a strategic reserve of low-enriched uranium to ensure 
uninterrupted operation of its light-water reactors'' 
(\underline{http://www.nci.org/03NCI/12/PL-Jeju-Process.htm\#\_edn3\#\_edn3}, 
ac\-ces\-sed January 31, 2004). Actual operating cost estimates for a 
facility still to be started up are obviously uncertain, but in any 
case it seems likely that the savings from mothballing the Rokkasho 
plant could nearly or completely pay for switching to constructing 
and operating ITER at Rokkasho instead. While some maintain that the 
primary if implicit purpose of reprocessing at Rokkasho is to provide 
latent nuclear weapons capability, a more immediate driver is the 
deal that was made with this remote prefecture that interim nuclear 
waste storage will be allowed at Rokkasho if and only if another 
major facility is built there as a quid pro quo economic benefit for 
the local economy. Under budgetary pressure an obvious alternative is 
to build ITER at Rokkasho \emph{instead} of completing and operating the 
reprocessing plant there. Although the alternative of promptly and 
completely abandoning building either could ideally be more cost 
effective, the chances of this being politically feasible for Japan 
appear very small.

Should the opportunity arise to \emph{substitute} ITER for operating the 
Rokkasho reprocessing plant arise, then the overall impact from an 
arms control, proliferation, and nuclear fuel cycle point of view 
could quite possibly be preferable compared to proceeding with 
reprocessing plant operations, particularly if the alternative ends 
up being also constructing ITER elsewhere with heavy Japanese 
participation (this being the only way ITER can be funded). Careful 
consideration should be given to the politics and technicalities of 
this alternative in any future work done on this subject. The details 
would depend on whether this involves abandoning reprocessing plant 
construction altogether or possibly operating it only on a pilot 
scale or even mothballing or dismantling it before it is heavily 
radioactively contaminated.

% Tables
% ======
\chapter{Tables}

\newpage\thispagestyle{empty}
%
% TABLE: Atomic and nuclear processes   File: processes.tex      25 April 1998
% ===================================                          28 January 2004
%
\begin{table}
\hskip 2cm \begin{tabular}{|l|l|l|} 		\hline
 & \raisebox{-0.8em}{{\bf atomic processes}} & \raisebox{-0.8em}{{\bf nuclear processes}}  \\
 & & \\ \hline
{\bf standard} & chemical detonation & fission \\
{\bf processes} & lasers & fusion \\ 
 &  & acceleration \\  \hline
{\bf advanced} & magnetic compression & subcritical fission \\
{\bf processes} & atomic isomerism & nuclear isomerism \\
 & x-ray lasers & $\gamma$-ray lasers \\
 & superlasers & muon catalysis \\ 
 & & antimatter \\
 \hline
{\bf exotic}    & metallic hydrogen & superheavy nuclei \\
{\bf processes} & atomic clusters  & bubble nuclei     \\
                & etc.              & halo nuclei       \\
 & & etc. \\ \hline
\end{tabular}
\caption{Major atomic and nuclear processes}
\hspace{2.0cm} of importance to ~present and future ~military explosives.    %\label{`tab:1'}
\end{table}
%

%--------------------------------------------------

\newpage\thispagestyle{empty}
%
% TABLE: LASER ICF FACILITIES   File: laser-ICF-7.tex   22 September 2000
%============================         laser-ICF.tex     28   January 2004
%
\begin{table}
\vspace*{-2cm} 	% vspace for report	
\begin{tabular}{|l|l|l|r@{}@{}l|r|r|l|}
\hline
\multicolumn{8}{|c|}{\raisebox{+0.4em}{{\bf \Large Laser beam driven ICF facilities \rule{0mm}{6mm}}}} \\ \hline
Country\raisebox{+1.em}{~} & System name & Location & \multicolumn{2}{|c|}{Energy} & No.~ & Wave &  \\ 
    &   &   & [kJ]/ & [ns] & $\!$beams$\!$ & length$\!\!\!\!$\raisebox{-0.5em}{~} &   \\   
\hline   
\multicolumn{8}{|c|}{{\bf Glass lasers}\raisebox{+1.em}{~}\raisebox{-0.5em}{~}} \\  
\hline
USA\raisebox{+1.em}{~} & Omega & LLE & 3/ & 0.6 & 24~ & 0.35~ &   \\
    & Omega-UG & LLE & 40/ & 3 & 60~ & 0.35~ & C \\
    & Nova & LLNL & 50/ & 1\hbox{~~} & 10~ & 0.35~ & T \\ 
    & NIF & LLNL & 1800/ & 5 & 192~ & 0.35~ & C \\  
\hline   
Japan\raisebox{+1.em}{~} & Gekko-XII & Osaka & 30/ & 1 & 12~ & 1.06~ &  \\ 
    & FIREX-II & Osaka &  50/ & 3 & 60$\sim$96~ & 0.35~ & D \\ 
%    & Kongoh & Osaka & 300/ & 3 & 92~ & 0.35~ & D \\ 
    & Koyo & Osaka & 4000/ & 6 & 400~ & 0.35~ & D \\    
\hline
France\raisebox{+1.em}{~} & LULI & Palaiseau  & 0.5/ & 0.6 & 6~ & 1.06~ &  \\ 
    & Octal & Limeil  & 0.9/ & 1 & 8~ & 1.06~ & ~ \\ 
    & Phébus & Limeil & 14/ & 2.5 & 2~ & 0.53~ & T \\    
    & Mégajoule & Bordeaux  & 1800/ & 15 & 288~ & 0.35~ & C \\   
\hline
China\raisebox{+1.em}{~} & Shen-Guang-I & Shanghai & 1.8/ & 1 & 2~ & 1.06~ &  \\
    & Shen-Guang-II & Shanghai & 6.4/ & 1 & 8~ & 1.06~ &  \\ 
    & Shen-Guang-III & Shanghai & 60/ & 1 & 60~ & 0.35~ & D \\   
\hline
UK\raisebox{+1.em}{~} & Helen & AWE & 1/ & 1 & 3~ & 0.53~ &  \\ 
    & Vulcan & RAL & 3/ & 1 & 6~ & 0.53~ & \\   
\hline
Russia\raisebox{+1.em}{~} & Delfin & Moscow & 3/ & 1 & 108~ & 1.06~ & \\  
\hline
India\raisebox{+1.em}{~} & ~ & Indore & 0.4/ & 3 & 4~ & 1.06~ & \\  
\hline
Italy\raisebox{+1.em}{~} & ABC & Frascati & 0.2/ & 2 & 2~ & 0.53~ &  \\   
\hline
Israel\raisebox{+1.em}{~} & ALADIN & Soreq & 0.1/ & 3 & 1~ & 1.06~ & \\  
    & Continuum & Soreq  & 0.07/ & 7 & 1~ & 1.06~ &   \\  
\hline
Germany\raisebox{+1.em}{~} &   & GSI  & 0.1/ & 15  & 1~ &     & \\  
              & PHELIX booster & GSI  & 4/   & 10  & 1~ &   &  C \\  
\hline
South Korea\raisebox{+1.em}{~} & Sinmyung-I & Taejon & 0.08/ & 0.5 & 1~ & 1.06~ &  \\   \hline
\multicolumn{8}{|c|}{{\bf KrF lasers}\raisebox{+1.em}{~}\raisebox{-0.5em}{~}} \\    
\hline
USA\raisebox{+1.em}{~} & Mercury & LANL & 1/ & 5 & 1~ & 0.25~ &  \\
    & Nike & NRL & 5/ & 4 & 56~ & 0.25~ & C \\    
\hline
Japan\raisebox{+1.em}{~} & Ashura & Ibaraki & 0.7/ & 15 & 6~ & 0.25~ & \\  
    & Super-Ashura& Ibaraki & 7/ & 22 & 12~ & 0.25~ & C \\   
\hline
UK\raisebox{+1.em}{~} & Sprite & RAL & 0.09/ & 60 & 6~ & 0.25~ & \\ 
    & Titania & RAL & 0.85/ & 0.5 & 1~ & 0.27~ &  \\   
\hline
China\raisebox{+1.em}{~} & Tin-Guang  & Shanghai & 0.4/ &   & 1~ & ~ & \\ \hline
\multicolumn{8}{|c|}{{\bf Iodine lasers}\raisebox{+1.em}{~}\raisebox{-0.5em}{~}} \\    
\hline
Russia\raisebox{+1.em}{~} & Iskra-5 & VNIIEP & 15/ & 0.25 & 12~ & 1.30~ &   \\   
\hline
Germany\raisebox{+1.em}{~} & Asterix IV & Garching & 1/ & 0.3 & 1~ & 1.30~ & T  \\
\hline   
Czech Republic\raisebox{+1.em}{~}  &    PALS    &  Prague  & 1.2/ & 0.3 & 1~ & 1.30~ &    \\   
\hline
\end{tabular}
\caption{Major operating, planned, or dismantled laser driven ICF facilities.} In the last column C means that the facility is under construction, D that it is in the design stage, and T that is has been taken to pieces. The wave length is in $\mu\rm m$.
\end{table}
%

%----------------------------------------------------

\newpage\thispagestyle{empty}
%
% TABLE: superlasers   File: superlaser-7.tex    20 september 2000
% ==================         superlaser-8.tex    28 January 2003
%
\begin{table}
\vspace*{-2cm}  % vspace for report
\begin{tabular}{|l|l|r|r|r|c|} 
\hline
\multicolumn{6}{|c|}{\raisebox{+0.4em}{{\bf \Large Superlasers \rule{0mm}{6mm}}}} \\ 
\hline
\raisebox{+0.2em}{Name}\rule{0mm}{6mm} & \raisebox{+0.2em}{Location} & \raisebox{+0.2em}{Energy} & \raisebox{+0.2em}{Duration} & \raisebox{+0.2em}{Power} & \raisebox{+0.2em}{Intensity}\\  
 & & \raisebox{+0.2em}{[J]}~~~ & \raisebox{+0.2em}{[ps]}\,~~ & \raisebox{+0.2em}{[TW]}\, & \raisebox{+0.2em}{[W/cm$^2$]}\\ 
\hline
\multicolumn{6}{|c|}{\raisebox{+0.2em}{\bf USA} \rule{0mm}{6mm}} \\ 
\hline
Petawatt\rule{0mm}{5mm}
         & LLNL(dismantled) & 1000.00 \quad &20-0.500 &    (1000.0)& $  >10^{21}$ \\
JanUSP   & LLNL           &   15.00 \quad &   0.085 &      200.0 & $2~~10^{21}$ \\
         & UM, Ann Arbor  &    3.00 \quad &   0.400 &        4.0 & $4~~10^{18}$ \\
Trident  & LANL           &    1.50 \quad &   0.300 &        5.0 & $  >10^{19}$ \\
         & UC, San Diego  &    1.00 \quad &   0.020 &       50.0 &              \\
LABS II  & LANL           &    0.25 \quad &   0.300 &$\sim$  1.0 & $1~~10^{19}$ \\
%        & UI, Chicago    &    0.15 \quad &   0.500 &        0.3 & $1~~10^{18}$ \\
         & UM, Ann Arbor  &    0.07 \quad &   0.025 &        3.0 &              \\
         & WSU, Pullman   &    0.06 \quad &   0.026 &        2.0 &              \\
%         & Stanford U.    &    0.06 \quad &   0.120 &        0.5 & $  >10^{18}$ \\
%
\hline
\multicolumn{6}{|c|}{\raisebox{+0.2em}{\bf Japan} \rule{0mm}{6mm}} \\ 
\hline
PW\rule{0mm}{5mm}
         & ILE            &  500.00 \quad &   0.500 &    1000.0  & $ >10^{20}$   \\
Petawatt & APRC (JAERI)   &   30.00 \quad &   0.030 &     850.0  & $ >10^{20}$  \\
PW-M     & ILE            &   60.00 \quad &   0.500 &     100.0  & $ >10^{19}$  \\
         & RIKEN          &    0.05 \quad &   0.500 & $\sim 0.1$ & $1~~10^{17}$ \\
\hline
\multicolumn{6}{|c|}{\raisebox{+0.2em}{\bf UK} \rule{0mm}{6mm}} \\ 
\hline
Vulcan\rule{0mm}{5mm}
         & RAL            &  360.00 \quad &   0.700 &      500.0 & $5~~10^{20}$ \\
%Titania & RAL            &   80.00 \quad &  60.000 & $\sim$ 1.3 &              \\
Astra    & RAL            &    1.00 \quad &   0.100 & $\sim 10.0$&   
\\
Titania  & RAL            &    1.00 \quad &   0.400 & $\sim  2.5$&              \\
Sprite   & RAL            &    0.25 \quad &   0.380 & $\sim  0.7$& $4~~10^{17}$ \\
\hline
\multicolumn{6}{|c|}{\raisebox{+0.2em}{\bf France} \rule{0mm}{6mm}} \\
\hline
Petawatt\rule{0mm}{5mm}
         & CESTA, Bordeaux& 1000.00 \quad &    1.000 &    1000.0 &     Constr.  \\
P-102    & CEL-V, Limeil  &   50.00 \quad &    0.500 &      80.0 & $  >10^{19}$ \\
         & LOA, Palaiseau &    0.80 \quad &    0.030 &      30.0 & $5~~10^{19}$ \\
         & LOA, Palaiseau &    0.03 \quad &    0.100 & $\sim 0.3$& $1~~10^{18}$ \\
ELIA     & U. of Bordeaux &    0.01 \quad &    0.010 &       1.0 & $1~~10^{18}$ \\
\hline
\multicolumn{6}{|c|}{\raisebox{+0.2em}{\bf Germany} \rule{0mm}{6mm}} \\ 
\hline
PHELIX  \rule{0mm}{5mm}
         & GSI, Darmstadt&  1300.00 \quad &    0.420 & $\sim 1000.0$&    Constr.  \\
Ti-Nd    & MBI, Berlin   &    10.00 \quad &    0.100 & $\sim  100.0$&    Constr.  \\
ATLAS    & MPQ, Garching &     5.00 \quad &    0.100 & $\sim  100.0$&    Constr.  \\
Ti       & MBI, Berlin   &     0.30 \quad &    0.032 & $\sim   10.0$& $ <10^{19}$ \\
ATLAS    & MPQ, Garching &     0.80 \quad &    0.130 & $\sim    5.0$& $ <10^{18}$ \\
         & IOQ, Jena     &     0.22 \quad &    0.110 & $\sim    2.2$& $ <10^{18}$ \\
\hline
\multicolumn{6}{|c|}{\raisebox{+0.2em}{\bf Russia} \rule{0mm}{6mm}} \\ 
\hline
Progress-P & St. Petersburg\rule{0mm}{5mm} & 55.00 \quad & 1.500 & $\sim 30.0$ & $1~~10^{19}$ \\
\hline
\multicolumn{6}{|c|}{\raisebox{+0.2em}{\bf China} \rule{0mm}{6mm}} \\ 
\hline
~        \rule{0mm}{5mm}
BM       &               &                &          &  $\sim 3.0$ &              \\
\hline
\end{tabular}
\caption{Major operating or planned superlaser facilities.} Interactions of \emph{super}lasers with matter are qualitatively very different from those of ordinary lasers because they have sufficient power (i.e., petawatt level) to induce strong relativistic, multi-photon, nonlinear, and direct nuclear (e.g., fission or fusion) effects.
%\label{`tab:xxt'}
\end{table}

% 

%-----------------------------------------------------

\newpage\thispagestyle{empty}
%
% TABLE: PARTICLE BEAM ICF SYSTEMS  File: particle-ICF-6.tex    11    Sept 1999
%=================================        particle-ICF.tex      28 january 2004
%
\begin{table}
\begin{tabular}{|l|l|l|r@{}@{}l|r|l|} 		
\hline
\multicolumn{7}{|c|}{\raisebox{+0.4em}{{\bf \Large Particle beam driven ICF facilities \rule{0mm}{6mm}}}} \\ 
\hline
Country\raisebox{+1.em}{~} & System name & Location & \multicolumn{2}{|c|}{Energy} & No.~  &  \\ 
    &   &   & [kJ]/ & [ns] & $\!$beams$\!$ \raisebox{-0.5em}{~} &   \\   
  
\hline 
USA\raisebox{+1.em}{~} & Saturn & SNL & 400/ & 5 & 36~  &   \\ 
                                  & PBFA--Z & SNL & 1500/ & 20 & 36~  & \\
                                  & ILSE & LBL & 6400/ & 10 &  16~  & D  \\   
\hline Germany\raisebox{+1.em}{~} & KALIF & Karlsruhe & 40/ & 40 & 1~  &  \\ 
    & HIBALL &   & 5000/ & 20 & 20~  & D \\
   
\hline Europe\raisebox{+1.em}{~} & HIDIF & ~ & 3000/ & 6 & 48~  & D \\ 
\hline
\end{tabular}
\caption{Major operating or planned particle-beam driven ICF facilities.}
\hspace{1.2cm} In the last column D means that the facility is in the design stage. 
\end{table}

%-----------------------------------------------------

\newpage\thispagestyle{empty}
%
% TABLE: Tritium breeding by fusion  File: breedtritium.tex      30 May 1998
% =================================                          28 January 2004
%

\begin{table}
\hskip 2cm \begin{tabular}{|l|lcl|}
\hline
              &                      &               & \\
              & {\bf Reactions}      & $\rightarrow$ & {\bf Products}  \\
              &                      &               & {\bf $-$ absorption plus leakage} \\
              &                      &               & {\bf $-$ consumption} \\
              &                      &               & {\bf $+$ heat} \\
              &                      &               & \\
\hline
              &                      &               & \\
{\bf Fusion}  & $D+T$                & $\rightarrow$ & $n+17.6$MeV \\
{\bf breeder} & $n+Be$               & $\rightarrow$ & $2.3 n - 0.7 n$  \\ 
              & $1.6 n + Li$         & $\rightarrow$ & $1.6 T - 1 T + 7.7$MeV \\
              &                      &               & \\
              & {\bf Total:}         &               & $0.6 T + 25$MeV \\
              &                      &               & \\
\hline
              &                      &               & \\
{\bf Fission} & $n + ~^{235}U$       & $\rightarrow$ & $2.5 n - 0.9 n - 1 n + 200$MeV \\
{\bf breeder} & $0.6 n + Li$         & $\rightarrow$ & $0.6 T + 3$MeV \\
              &                      &               & \\
              & {\bf Total:}         &               & $0.6 T + 200$MeV \\
              &                      &               & \\
\hline
\end{tabular}
\caption{Fusion and fission tritium breeding reactions in dedicated facilities.} \emph{Adapted from W.A. Lokke and T.K. Fowler,  Report UCRL--94003, Lawrence Livermore National Laboratory (1986)}.
\end{table}
%

%-----------------------------------------------------

% Figures  (Size set by height=23 or width=16)
% =======
\chapter{Figures}

\pagestyle{empty}

\newpage\thispagestyle{empty}
\begin{figure}
\resizebox{16cm}{!}{ \includegraphics{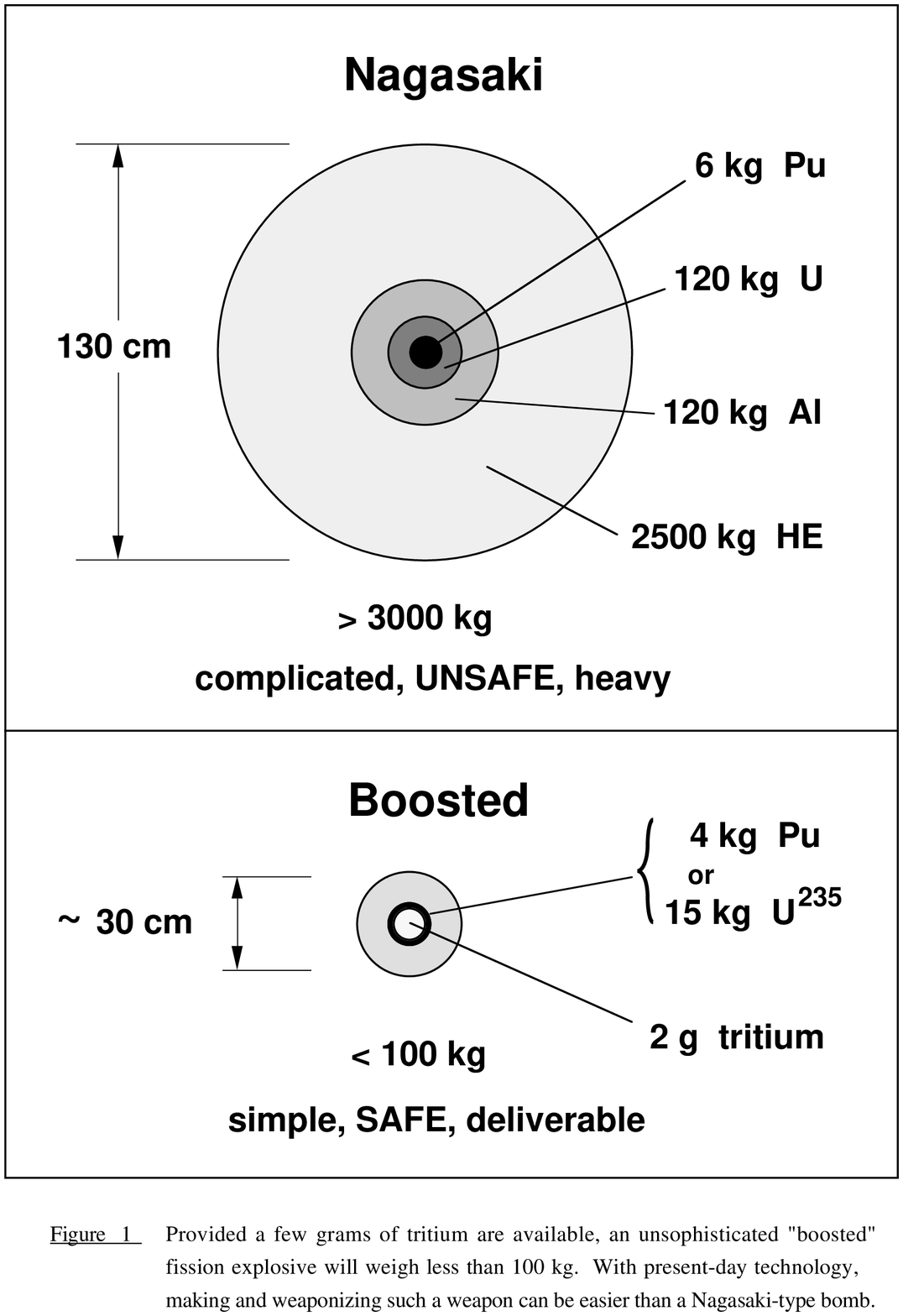}}
%------------------------------------------------------------
\caption{Comparison of Nagasaki-type  and tritium-boosted A-bombs}
\end{figure}

\newpage\thispagestyle{empty}
\begin{figure}
\resizebox{16cm}{!}{ \includegraphics{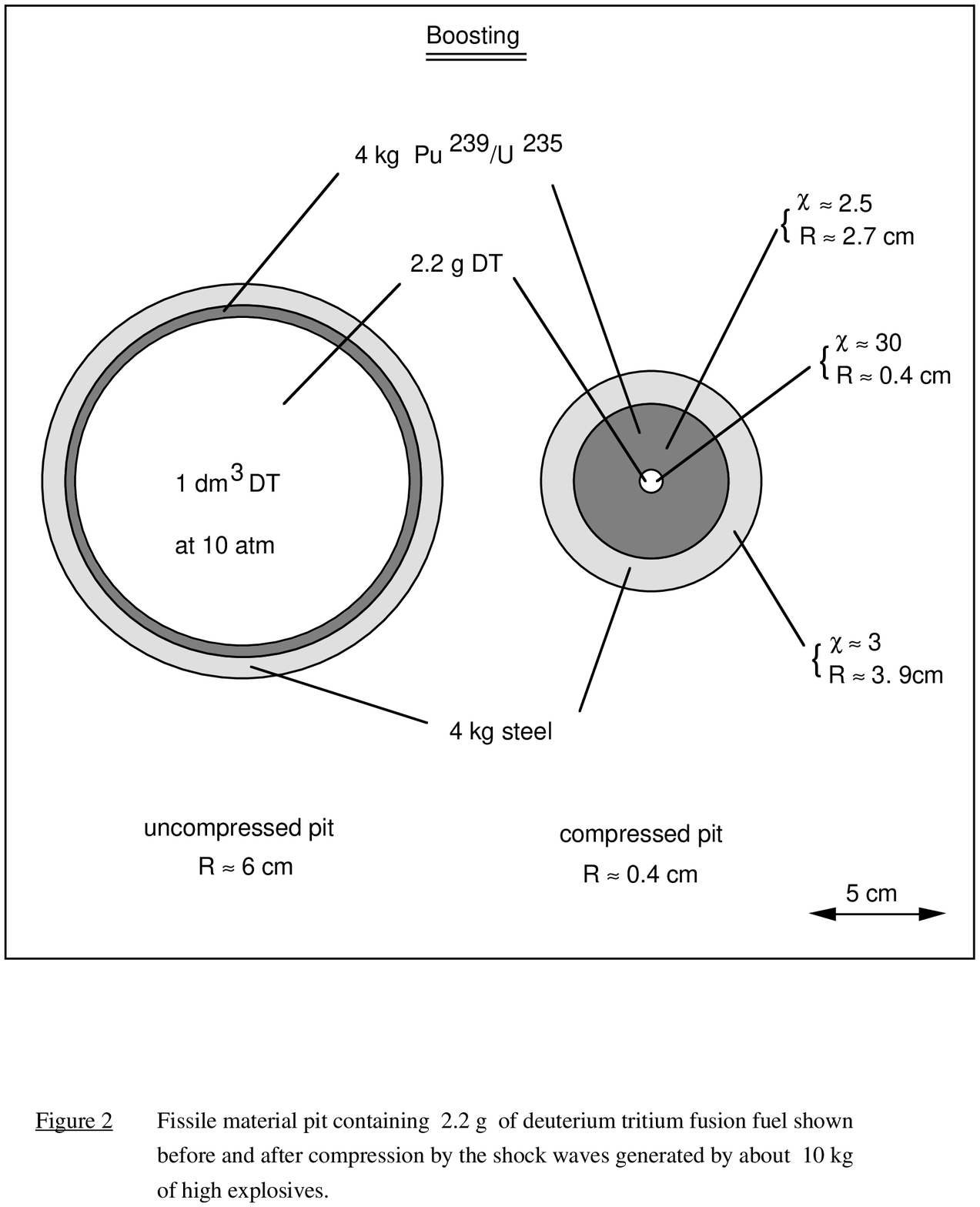}}
%------------------------------------------------------
\caption{Principle of tritium-boosted A-bombs}
\end{figure}

\newpage\thispagestyle{empty}
\begin{figure}
\resizebox{16cm}{!}{ \includegraphics{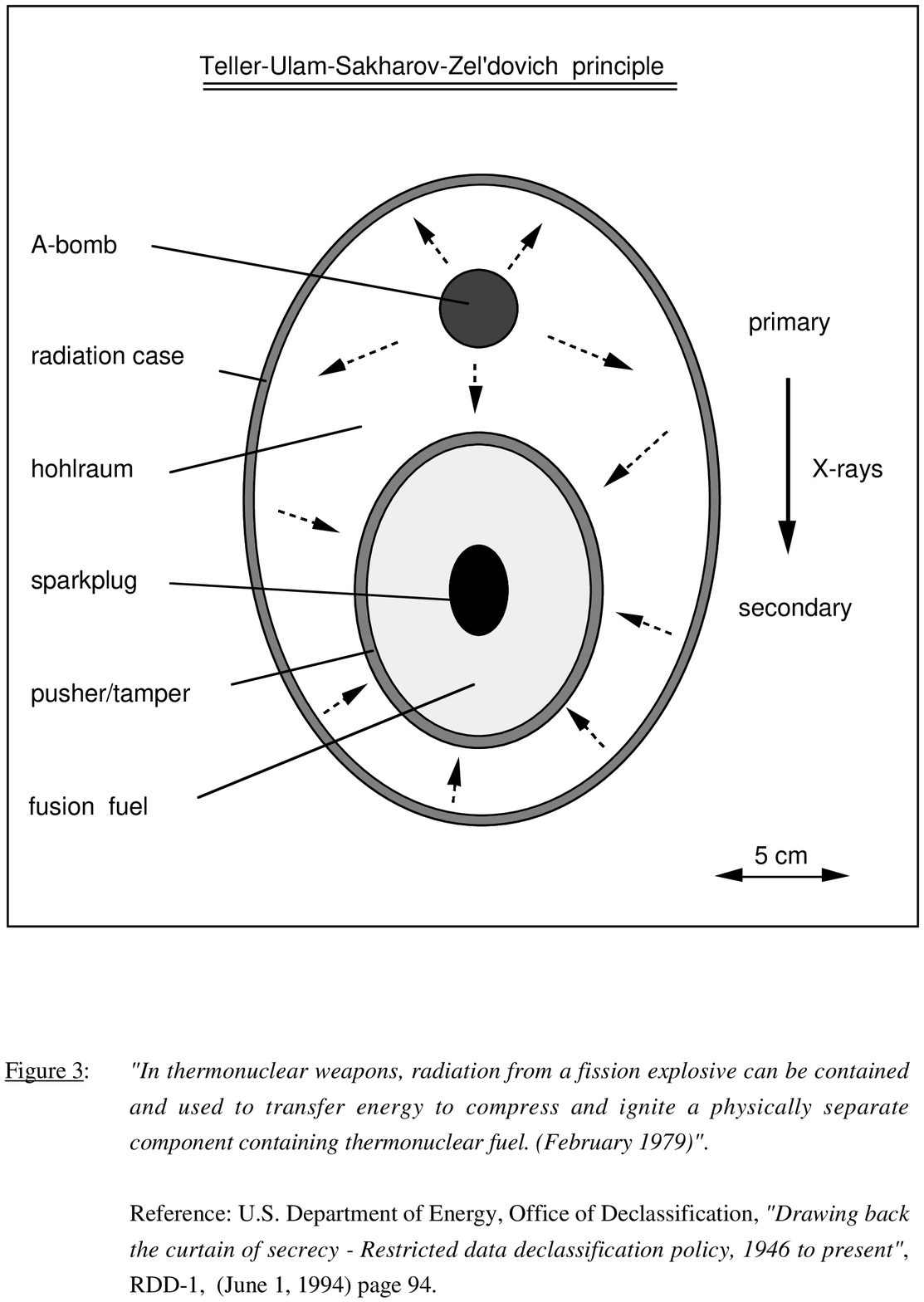}}
%-------------------------------------------------------
\caption{Teller-Ulam principle of two-stage H-bombs}
\end{figure}

\newpage\thispagestyle{empty}
\begin{figure}
\resizebox{16cm}{!}{ \includegraphics{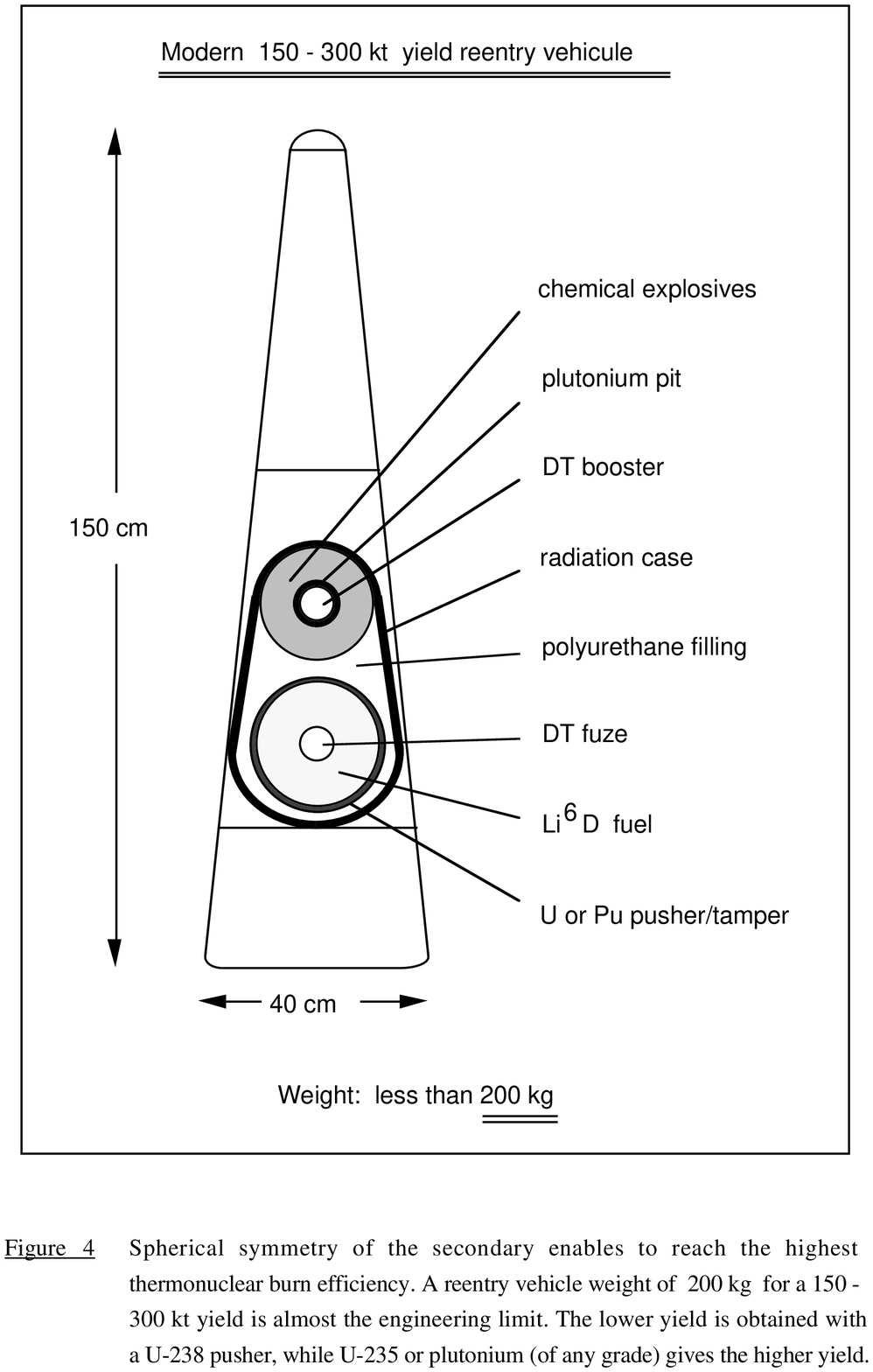}}
%-------------------------------------------------------
\caption{Modern 150 - 300 kt yield reentry vehicle}
\end{figure}

\newpage\thispagestyle{empty}
\begin{figure}
\resizebox{16cm}{!}{ \includegraphics{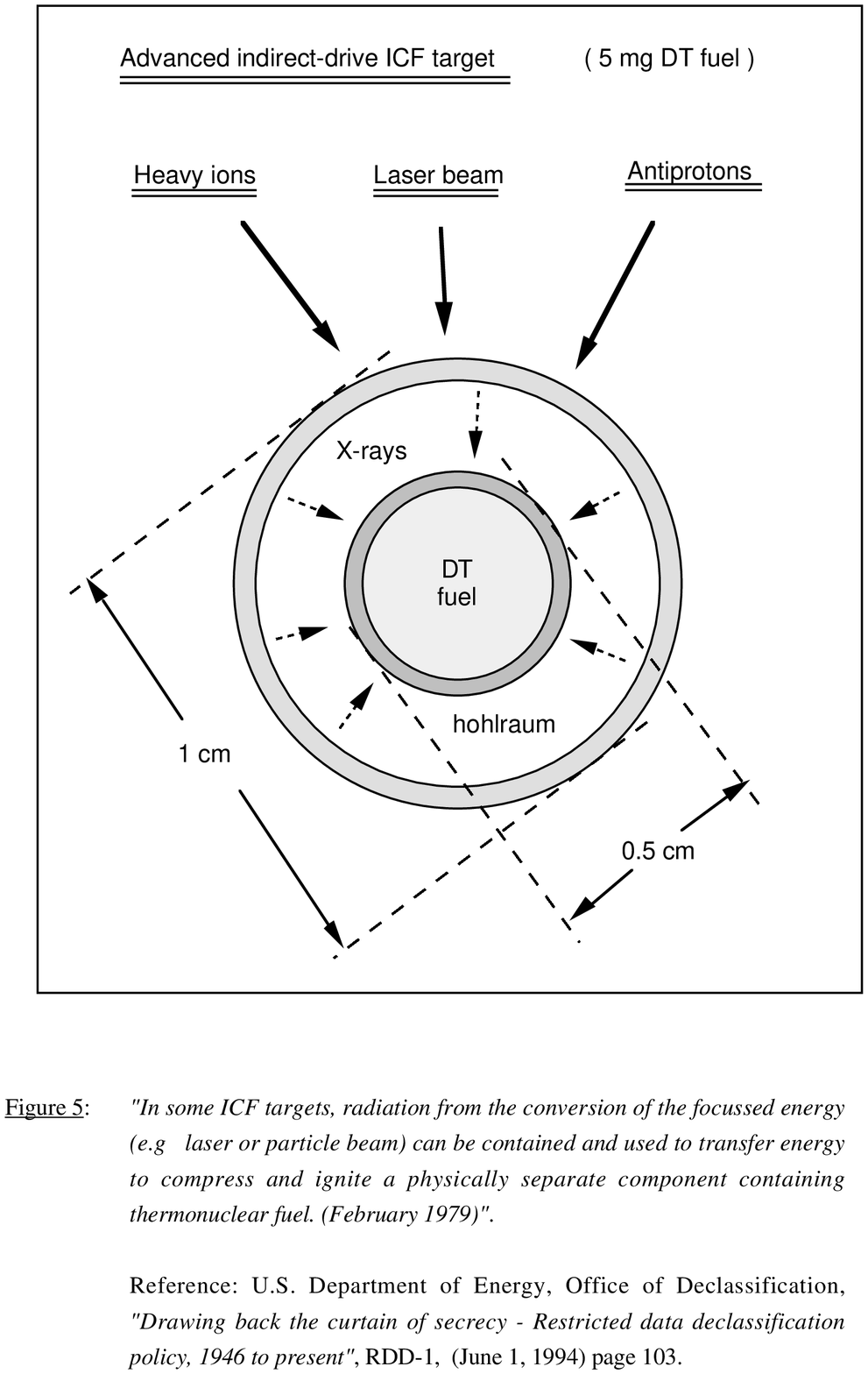}}
%-------------------------------------------------------
\caption{Advanced indirect-drive ICF target}
\end{figure}

\newpage\thispagestyle{empty}
\begin{figure}
\resizebox{16cm}{!}{ \includegraphics{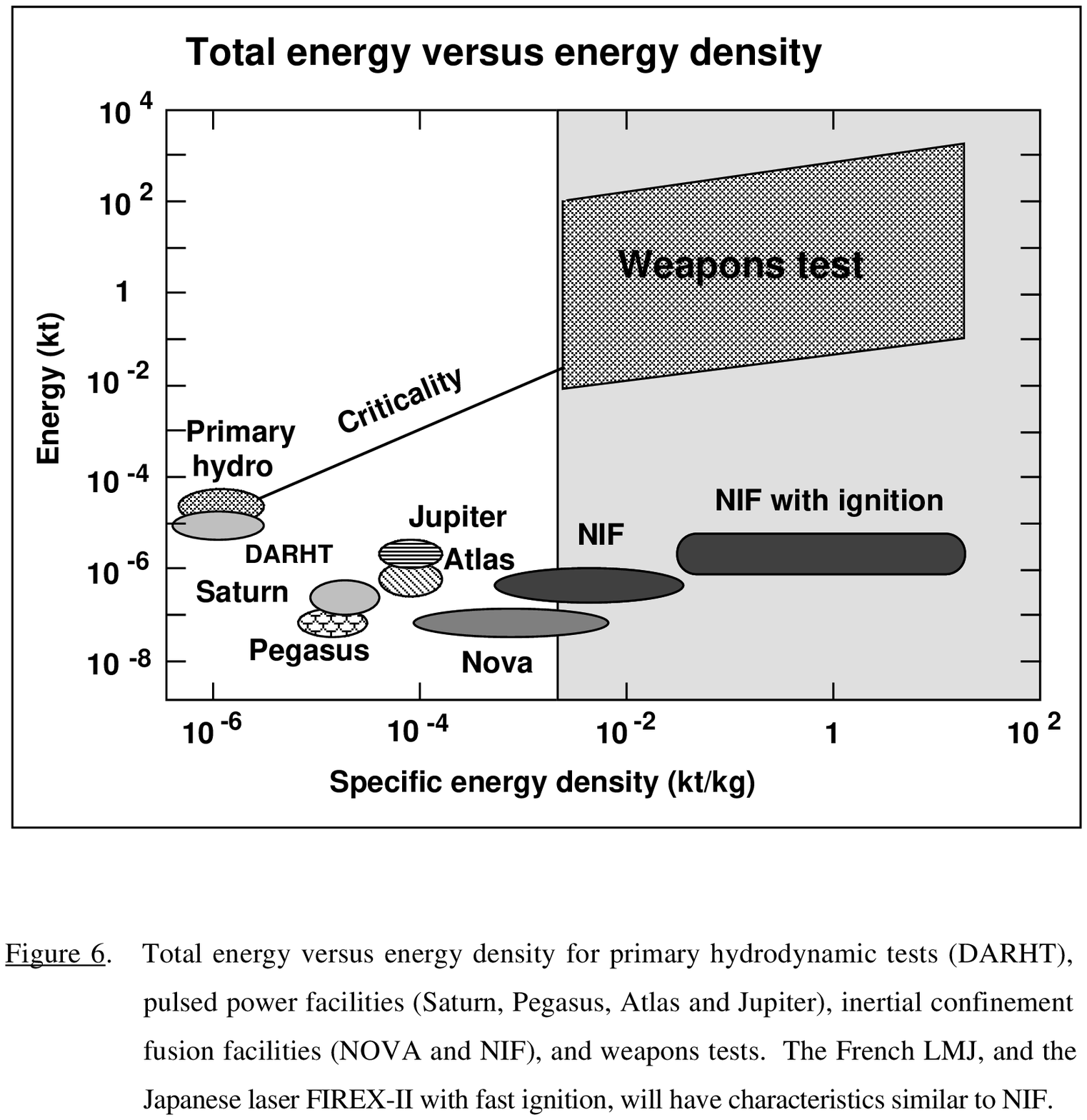}}
%-------------------------------------------------------
\caption{Total energy versus energy density}
\end{figure}

\end{document}